# Projective geometry and special relativity

**D.H. Delphenich***

Physics Department, Bethany College, Lindsborg, KS 67456, USA

Some concepts of real and complex projective geometry are applied to the fundamental physical notions that relate to Minkowski space and the Lorentz group. In particular, it is shown that the transition from an infinite speed of propagation for light waves to a finite one entails the replacement of a hyperplane at infinity with a light cone and the replacement of an affine hyperplane – or rest space – with a proper time hyperboloid. The transition from the metric theory of electromagnetism to the pre-metric theory is discussed in the context of complex projective geometry, and ultimately, it is proposed that the geometrical issues are more general than electromagnetism, namely, they pertain to the transition from point mechanics to wave mechanics.

**Contents**

* delphenichd@bethanylb.edu

## 1 Introduction

In all of the time that has elapsed since Einstein first started constructing the geometric basis for gravitational motion in the universe, one thing has been largely accepted without question. Except for the predictable attempts at going beyond the scope of general relativity – e.g., Kaluza-Klein theory, projective relativity, strings, and what have you – the divinity of the Minkowski space scalar product and the associated Lorentz group of transformations has been widely regarded as the least questionable foundation of spacetime structure.

However, starting with the work of Cartan [**1**], Kottler [**2**], and Van Dantzig [**3**] on pre-metric electromagnetism, which was later expanded and refined in various ways by Post [**4**], Hehl and Obukhov [**5**], and others, a picture began emerging that would definitively challenge that article of faith. The basic gist of the argument is this: The scalar product on Minkowski space is actually an abstraction of something that is more fundamentally related to the way that spacetime supports the propagation of electromagnetic waves, namely, the electromagnetic constitutive law of spacetime. In the linear case, such an electromagnetic constitutive law defines an almost-complex structure on the bundle of 2-forms on spacetime. When that linear electromagnetic constitutive law is also symmetric, it defines a complex orthogonal structure on the fibers of that bundle, and the associated structure group is then $SO(3; \mathbb{C})$, which is isomorphic to the connected component of the Lorentz group. Hence, one should shift one's attention from describing the geometrical structure of spacetime as described by the Riemannian picture of tangent vectors and Lorentzian metrics to describing it in terms of bivectors or 2-forms and an almost-complex structure on the bundle in question. This suggests a more Kleinian picture of complex projective geometry at the foundations of spacetime structure.

The primary purpose of this study is to first establish that the introduction of projective geometry into relativistic physics is natural and appropriate even in the context of Minkowski space, even if only to account for the introduction of a conformal Lorentz structure, and ultimately, a Lorentz structure. We shall then present a brief synopsis of pre-metric electromagnetism and discuss how many of the concepts of special relativity can be represented in terms of bivectors or 2-forms instead of tangent vectors. Finally, we shall point out that along with the shift from tangent vectors to bivectors, we are really also making a shift from point mechanics to wave mechanics, so discussing the geometrical structure of spacetime by means of the bundle of 2-forms may also have the advantage of carrying with it a more quantum-physical context, as well.

## 2 Projection and measurement

One way of going beyond the usual metric formulation of electromagnetism is to observe that some of the most elementary processes in special relativity can be given straightforward explanations using the language of projective geometry ([1]) as their basis, and then identifying the conditions that bring about the reductions to affine, conformal, and metric geometry. The picture that emerges is that when $\mathbb{R}^4$ is regarded as the space of homogeneous coordinates for the various three-dimensional affine subspaces of $\mathbb{R}^4$ that one obtains by projection from the origin, one finds that it is really the projective geometry of these affine subspaces that is being described by the methods of special relativity.

[1] Although the formulation of the relevant notions of projective geometry in this article is somewhat specialized to the demands of spacetime structure, nevertheless, it does not diverge appreciably from what one can find in various texts on analytical projective geometry [**6-9**].

### 2.1 Rest spaces, hyperplanes at infinity, proper time lines

The physical concept that underlies all of this is that physical measurements seem to address only the projections of physical observables into the rest space of the measuring devices. Now, the notion that a measurement is a projection of a state space onto a subspace is well-established in the statistical interpretation of quantum wave mechanics. Nevertheless, the same concept is rarely emphasized in the context of spacetime structure itself.

In order to relate the concept of rest space for a motion to a 1+3 decomposition of spacetime, we define a *motion* in a differentiable (configuration) manifold $M$ as a map $x: T \times \mathcal{I} \to M$, $(\tau, \alpha) \mapsto x_\alpha(\tau)$, in which the subset $T \subset \mathbb{R}$ represents the proper time parameter and the set $\mathcal{I}$ indexes the objects that are moving. The set $\mathcal{I}$ can be finite, as in the case of planets, colliding masses, or even the molecules in a volume of a gas, or uncountable, as in the description of continuous matter. In the finite case a motion is then a finite set of curves in $M$, and in the uncountable case it usually takes the form of a congruence of curves, such as one might obtain from the flow of a vector field, where it exists. Hence, for the present purposes, we will be glossing over the subtleties that arise in the form of integrability issues.

We also assume that the map $x$ is differentiable for every value of $\alpha \in \mathcal{I}$. The *velocity vector field* of the motion is then the map $\mathbf{v}: T \times \mathcal{I} \to M$, $(\tau, \alpha) \mapsto \mathbf{v}_\alpha(\tau)$ that one obtains by differentiating with respect to the proper time parameter. Hence, the support of the velocity vector field is along the curves of the motion.

For a given object – say, $O \in \mathcal{I}$ – the *instantaneous rest space* of $O$ at the proper time instant $\tau \in T$ is the set of all $\alpha \in \mathcal{I}$ such that the velocity vector $\mathbf{v}_\alpha(\tau)$ "equals" $\mathbf{v}_O(\tau)$. Although in order to do justice to the concept of the equality of vectors tangent to different points of a manifold, in the present analysis, we shall mostly be concerned with $\mathbb{R}^4$, so we assume that $M$ is an open subspace of a vector space. This relationship defines an equivalence relationship on $\mathcal{I}$ whose classes are each instantaneous rest spaces for different velocities. Naturally, these classes may very well consist of single points in many cases. We could also characterize the elements of a rest space as all having zero relative velocity. A set of objects is called *co-moving* when they are all in the same rest space for all $\tau$.

Since we are assuming that $M$ is contained in a vector space $V$, we can also speak of the vector subspace $\Pi_\tau$ of $V$ that is spanned by the elements of an instantaneous rest space at the proper time instant $\tau$. If $V$ is of dimension $n$ then the dimension of $\Pi_\tau$ can be anywhere from 0 to $n - 1$, since no non-zero $\mathbf{v}_\alpha(\tau)$ can be an element of $\Pi_\tau$. Hence, in the maximal case, since all of the $\mathbf{v}_\alpha(\tau)$ are the same vector $\mathbf{v}(\tau)$, which we assume is also non-zero, the line $[\mathbf{v}(\tau)]$ that is generated by all scalar multiples of $\mathbf{v}(\tau)$, which we call the *proper time line at the instant* $\tau$, and the $n$–1-plane $\Pi_\tau$ collectively define a decomposition of $V$ into a direct sum $[\mathbf{v}(\tau)] \oplus \Pi_\tau$. Hence, any vector $\mathbf{w} \in V$ can be expressed uniquely as a sum of a vector along $[\mathbf{v}(\tau)]$ and a vector in $\Pi_\tau$, and these respective vectors are the projections of **w** into the respective subspaces. In particular, all of the velocity vectors for a given rest space will project to zero in $\Pi_\tau$.

The essential idea in physics to consider is, as we suggested above, that *any physical measurement is made in the rest space of the measuring device.* For instance, a measurement of a time interval by a chronometer will always represent a proper time interval in the rest space of the chronometer. Hence, one can think of the process of measurement as being fundamentally linked with the process of projecting from spacetime – which we regard as simply a four-dimensional affine space $V$, for the present purposes – onto a three-dimensional rest space $\Pi_3$ that is defined by the measuring device. Furthermore, the projection also depends upon one's choice of point in spacetime that would represent an observer or center of projection. Hence, any measurement implicitly relates to a choice of two geometric objects: a three-dimensional affine subspace $\Pi_3$ to serve as rest space and a point $\mathbb{O}$ that is not incident with $\Pi_3$ to serve as center of projection.

With such a choice of point and affine hyperplane, the points of the rest space $\Pi_3$ are then in one-to-one correspondence with the lines through $\mathbb{O}$, with the exception of the lines that lie in a linear hyperplane $\Pi_\infty$ that is incident with $\mathbb{O}$ and parallel to $\Pi_3$. One calls this hyperplane the *hyperplane at infinity.* Not coincidentally, it plays a role that is similar to that of the "celestial sphere" of astronomy. The latter geometric object lives in $\Pi_3$ and its pairs of antipodal points define the projectivization of $\Pi_3$, which is projectively equivalent to $\mathbb{R}P^2$; the set of lines through $\mathbb{O}$ is projectively equivalent to $\mathbb{R}P^3$.

Now, it is crucial to note that the projection in question at the moment – viz., $V \to \Pi_3$ – does not take the form of the projection of a Cartesian product of spaces onto a factor, which geometrically represents essentially an orthogonal projection – viz., a projection about a point at infinity – but a projection about a finite point. The difference between them is depicted in Fig. 1.

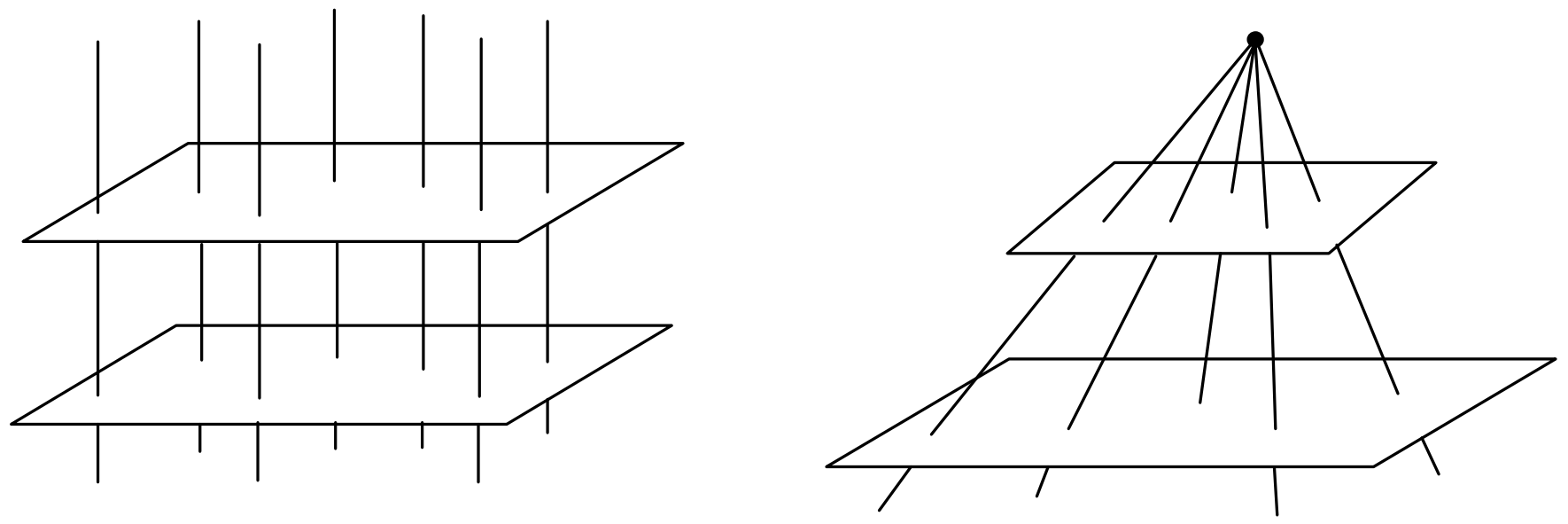

Fig. 1. Orthogonal projection vs. projection from a finite point.

Physically, this reduction from infinity to finitude corresponds to the reduction of $c$ from an infinite value, which gives the Galilei group, to a finite value, which gives the Lorentz group. This, of course, suggests that the location of the center of projection is related to the way that waves propagate in spacetime. We shall return to discuss this notion in more detail shortly, but first, we return to the nature of the hyperplane at infinity.

One can intuitively think of the hyperplane at infinity $\Pi_\infty$, which passes through $\mathbb{O}$, as being obtained by an inversion through the affine hyperplane $\Pi_3$ that takes $\infty$ to $1/\infty = 0$. In this picture, the hyperplane at infinity, like the celestial sphere, represents a sort of mythical geometrical object that is defined by the "positive endpoints" of the lines through $\mathbb{O}$. We express this schematically in Fig. 2:

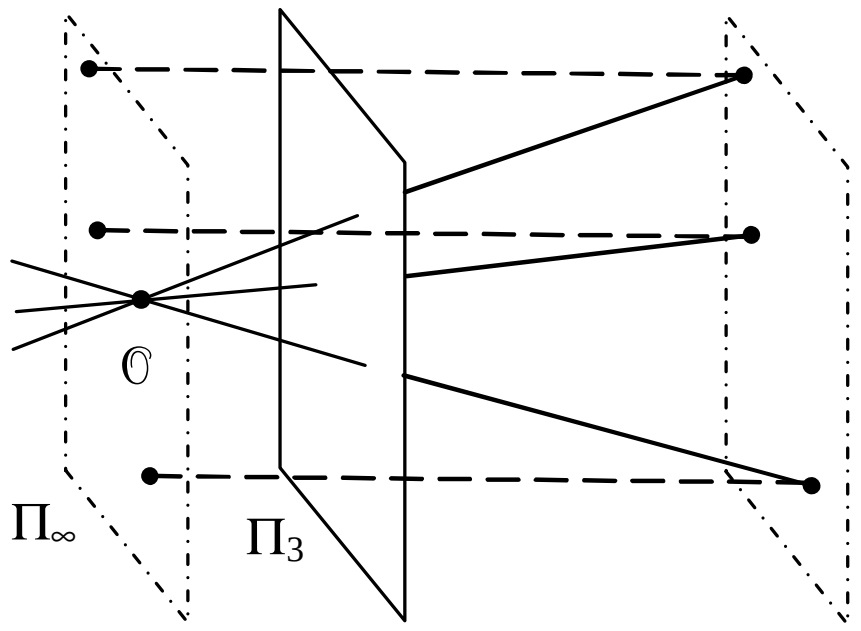

Fig. 2. Inversion of the hyperplane at infinity.

In keeping with the established language of relativity, we also refer to any line through $\mathbb{O}$ that intersects $\Pi_3$ as a *proper time line* (relative to $\mathbb{O}$ and $\Pi_3$), although, of course, in the absence of a Minkowski scalar product on $V$ we cannot definitively say if the line is actually timelike, lightlike, or spacelike. Once again, this suggests that we are trying to do physics in a spacetime in which electromagnetic waves propagate with infinite velocity, so we see that the projective geometry of lines through $\mathbb{O}$ and $\Pi_3$ is a sort of "static" limit of relativistic kinematics, in which the hyperplane at infinity can only be reached by lines that represent motions with infinite velocity.

One immediately notes that making a choice of proper time line $L$ is equivalent to making a choice of *spatial origin*, namely, the point in $\Pi_3$ that is incident on $L$. Conversely, if one has made a choice of spatial origin then making a choice of $L$ is *not* equivalent to making a choice of $\mathbb{O}$, but to making a choice of hyperplane at infinity, whose intersection with $L$ then gives a choice of observer $\mathbb{O}$.

For later reference, we introduce the notion of the projective space $PV_{\mathbb{O}}$, which is the manifold of all lines in $V$ that pass through $\mathbb{O}$. One can also regard $PV_{\mathbb{O}}$ as the projectivization of the tangent space $T_{\mathbb{O}}(V)$, i.e., the set of all lines through the origin in $T_{\mathbb{O}}(V)$. $PV_{\mathbb{O}}$ is non-canonically projectively equivalent to $\mathbb{R}P^3$, since one can define such an equivalence by any choice of frame on $V$, which amounts to an affine isomorphism of $V$ with $\mathbb{R}^4$, and would therefore take lines through $\mathbb{O}$ to lines through the origin of $\mathbb{R}^4$.

### 2.2 Introduction of a metric into a rest space and an arc-length on a proper time line

Of course, so far we have said nothing about the introduction of distance or coordinates into the picture that we have described, since distance defines a less general geometry than projective geometry. However, we eventually have to explain the transition from this pre-metric projective geometry to the indefinite metric of Minkowski space, so we should look at how one introduces the various pieces of that object one at a time.

The first reduction is from projective geometry to conformal geometry, which is effected by the introduction of a Euclidian scalar product $\delta$ on $\Pi_3$. (Strictly speaking, one introduces $\delta$ on the tangent spaces to $\Pi_3$, since the *points* of $\Pi_3$ are not assumed to be described by *vectors*, at least initially.) Equivalently, one can introduce a Euclidian quadratic form $Q_\delta$ on $\Pi_3$. The formulas that give each in terms of the other are:

$$Q_\delta(\mathbf{v}) = \delta(\mathbf{v}, \mathbf{v}), \tag{2.1a}$$
$$\delta(\mathbf{v}, \mathbf{w}) = \tfrac{1}{2}[\, Q_\delta(\mathbf{v}) + Q_\delta(\mathbf{w}) - Q_\delta(\mathbf{v} - \mathbf{w})] . \tag{2.1b}$$

One sees that a choice of $\delta$ on $\Pi_3$ is equivalent to defining an angle measure between the lines of $PV_{\mathbb{O}}$. However, the relationship is not as straightforward as simply saying that the angle between two lines in $PV_{\mathbb{O}}$ is proportional to the Euclidian distance between the points at which they intersect $\Pi_3$, since such a definition would not agree with the usual one, which relates to the distance between points on a sphere, not points in a 3-plane. One can see that the conventional Euclidian angle between two lines through $\mathbb{O}$ gets smaller for the same separation distance in $\Pi_3$ as $\Pi_3$ approaches the hyperplane at infinity.

A choice of $\delta$, in conjunction with a choice of $L$, defines a *unit 2-sphere* about the spatial origin in $\Pi_3$ that $L$ defines, and in the obvious way. This, in turn, is equivalent to defining a choice of spherical cone in $PV_{\mathbb{O}}$ with its vertex at $\mathbb{O}$ and its symmetry axis along $L$. Hence, we see that a choice of Euclidian scalar product on $\Pi_3$ is indeed equivalent to a conformal structure on $V$, which is either expressed as a choice of spherical cone in $V$ or a choice of angle measure on $PV_{\mathbb{O}}$.

In addition to measuring distances in our chosen rest space $\Pi_3$, it would also be convenient to have a way of measuring the lengths of intervals along a chosen $L$. This is equivalent to defining a parameterization of $L$ – viz., a diffeomorphism $\tau: L \to \mathbb{R}$ – since choosing such a parameterization implies that one can define the proper-time arc-length between two points $A$ and $B$ of $L$ to be:

$$d(A, B) = \tau(B) - \tau(A), \tag{2.2}$$

and conversely, if $d(A, B)$ is a choice of arc-length, then any choice of proper time origin on $L$ – i.e., any point $A$ of $L$ – defines a parameterization by way of

$$\tau(B) = d(A, B). \tag{2.3}$$

Although the parameterization of lines is usually treated by differential geometry as a mere convenience, that is because when one is dealing with a manifold that has a Riemannian (pseudo-Riemannian, resp.) structure, one can always (almost always, resp.) define a unique choice of parameterization by requiring that it have unit speed. In the absence of such a structure, there are actually certain subtleties regarding the nature of parameterizations that become more significant in projective geometry than in metric geometry. We shall discuss these issues in more detail below.

### 2.3 Homogeneous and inhomogeneous coordinates

Although the *concept* of a coordinate system as a homeomorphism of an open subset of topological space with $\mathbb{R}^n$ is sufficiently elementary as to seem unworthy of closer scrutiny, in fact, the *process* of actually defining one, as in physical applications, is really more involved than it would first appear. Indeed, when one tries to give a projective-geometric context, one sees that there are some subtleties that might prove significant. One sees that the process of defining what we shall call an *elementary coordinate system* seems to involve, at the least, the following machinery:

*i*) A distinguished point $\mathbb{O}$ to serve as origin.

*ii*) A minimal "spanning" set of lines $\{L_n, i = 0, \ldots, n\}$ that one usually calls *axes* (which might be curved geodesics). Collectively, these lines define a *projective frame.*

*iii*) A parameterization for each line; i.e., a diffeomorphism $\sigma_n$: $L_n \to \mathbb{R}$ for each $n$.

*iv*) A projection from each point of the space onto each line of the set.

*v*) Some sort of homogeneity to the space, such as the action of translations.

One such process in an $n$+1-dimensional affine space $V$ might involve the following recursion:

*i*) Choose a point $\mathbb{O}$ and a line $L_0$ through $\mathbb{O}$.

*ii*) Choose an $n$-plane $\Pi_n$ that includes the given point $x \in V$, but not $\mathbb{O}$ or $L_0$.

*iii*) Intersect the line $L_0$ with $\Pi_n$ at a point $\mathbb{O}'$.

*iv*) Call $x^0$ the distance from $\mathbb{O}$ to $\mathbb{O}'$, as measured by the parameterization $\sigma_0$.

*v*) Choose an $n$–1-plane $\Pi_{n-1}$ in $\Pi_n$ that includes the given point $x \in V$.

*vi*) Intersect the line $L_1$ through $\mathbb{O}'$ with $\Pi_{n-1}$ at a point $\mathbb{O}''$.

*vii*) Call $x^1$ the distance from $\mathbb{O}'$ to $\mathbb{O}''$, as measured by the parameterization $\sigma_1$.

*viii*) Repeat as necessary.

As long as $V$ is finite-dimensional, the recursion converges after a finite number of steps and produces a set of $n$+1 real numbers ($x^0, \ldots, x^n$) that uniquely determine $x$; i.e., $n$+1 real *coordinates*. Note that the assumption of the homogeneity of $V$ under translations is essential in order for the successive subspaces to be foliated by the next lower dimensional subspaces with codimension one so that one could always find a subspace that includes $x$ by simply varying the value of $x^k$.

The essential aspects of the aforementioned process for the moment are the fact that one must make a choice of projective frame and the fact that one is assuming that every line in the affine space $A$ has a chosen diffeomorphism with $\mathbb{R}$. Since all of the other parameterizations of a given line will be in one-to-one correspondence with the elements of the group of diffeomorphisms of $\mathbb{R}$, one can think of a choice of assignment as essentially a choice of a Diff($\mathbb{R}$) "gauge" for the projectivization of $V$ – viz., the set $PV$ of all lines in $V$ – which is also the projectivization PT($V$) of the tangent bundle to $V$, since a line through any point $x \in V$ will be uniquely associated with a line in $T_x(V)$.

If we return to the case at hand, of a four-dimensional affine space $V$ with a given choice of observer $\mathbb{O}$, proper time line $L_0$, and rest space $\Pi_3$ then we see that the process that we described is essentially the construction of the Cartesian coordinates of a point $x \in \Pi_3$, or the *homogeneous coordinates* ($x^0, \ldots, x^3$) of the line through $\mathbb{O}$ and $x$, which defines an element of $PT_{\mathbb{O}}(V)$, and which is, in turn, diffeomorphic to $\mathbb{R}P^3$. Although the Cartesian projection of $(x^0, \ldots, x^3)$ onto $(x^1, x^2, x^3)$ will give the Cartesian coordinates of the point $x$ in the vector space $\Pi_3$, nevertheless, in order to define a coordinate chart – called a *Plücker chart* – for the projective space $PT_{\mathbb{O}}(V)$, one has to introduce *inhomogeneous* coordinates for the line through $\mathbb{O}$ and $x$ by way of:

$$X^i = \frac{x^i}{x^0}, \quad i = 1, 2, 3 . \tag{2.4}$$

Actually, the chart thus defined does not cover all of $PT_{\mathbb{O}}(V)$ since it will miss the points of the hyperplane at infinity, for which $x^0 = 0$. One must then define other similar charts in each of the other spanning directions through $\mathbb{O}$, which then corresponds to dividing by $x^k$, $k = 1, 2, 3$, instead of $x^0$ in (2.4).

One immediately notes that the Cartesian projection will coincide with the projection of $(x^0, \ldots, x^3)$ onto $(X^1, X^2, X^3)$ in the event that $x^0 = 1$, so the only way that the two projection will differ is when $x^0 \neq 1$, which then depends upon the parameterization of $L_0$. Note that this implies that making a different choice of rest space $\Pi_3$ is, in a sense, equivalent to making a different choice of parameterization for the proper time line $L_0$. What the inhomogeneous coordinates of a line through $\mathbb{O}$ then give one is independence from the choice of either $\Pi_3$ or, equivalently, parameterization of $L_0$.

The introduction of a Lorentzian conformal structure into $V$ can be viewed in projective geometric terms as being equivalent to the introduction of a Euclidian scalar product (more precisely, angle measure) into $\mathbb{R}P^3$ by way of inhomogeneous coordinates, since the equation for the unit sphere in $\mathbb{R}^3$:

$$1 = \delta_{ij} X^i X^j = \delta_{ij} \frac{x^i}{x^0}\frac{x^j}{x^0} \tag{2.5}$$

is clearly equivalent to the equation for the cone in $\mathbb{R}^4$, as described in homogeneous coordinates:

$$0 = (x^0)^2 - \delta_{ij} x^i x^j . \tag{2.6}$$

Since our basic thesis here is that a physical measurement is something that can only involve the data that is obtained in $\Pi_3$ that data will be parameterized by *inhomogeneous* coordinates, not homogeneous coordinates. We shall next see that this is indeed consistent with the usual relativistic process of projecting tangent vectors from Minkowski space to a three-dimensional rest space.

### 2.4 The transition from c = ∞ to c = finite

We still need to establish what we mentioned above: that one way of characterizing the difference between classical mechanics and relativistic mechanics is in the difference between orthogonal projection – i.e., projection from a point at infinity – and projection from a finite point, and that this is physically related to the finitude of the speed of light.

What we shall see is that when $c$ goes from infinity to a finite value, the time coordinate $t$ in our four-dimensional affine space $V$ goes to the proper time parameter $\tau$ by way of a rescaling that differs for each point in $V$, the hyperplane at infinity goes to the light cone of Minkowski space, and the affine hyperplane $\Pi_3$, which is defined by $t = 1$ goes to the $\tau = 1$ hyperboloid. We illustrate this in Fig. 3, in which we have abbreviated the three space dimensions to $r$ by spherical symmetry:

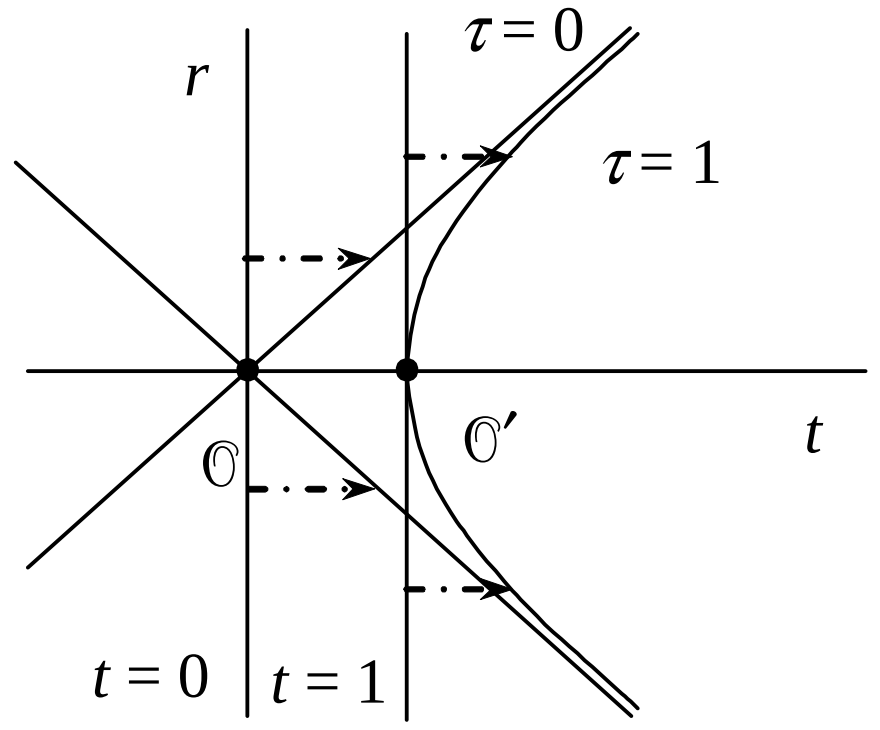

Fig. 3. The transition from $c$ infinite to $c$ finite.

It is important to note that the only effect on the lines through $\mathbb{O}$ amounts to a linear shearing transformation, which is still a projective transformation, since any linear transformation takes lines to lines.

Once again, assume that we have a Euclidian scalar product $\delta$ in our affine hyperplane $\Pi_3$, which we assume is defined by all points of $V$ that have homogeneous coordinates with $x^0 = 1$. We assume that the speed of light $c$ can vary from 0 to $+\infty$ and we can define the *light sphere* $S_c$ in $\Pi_3$ to be the 2-sphere centered at $\mathbb{O}'$, which is the intersection of the $x^0$-axis with $\Pi_3$ and defines the origin of the chosen rest space, and of radius $c$. Physically, this means that it will have a radius of one light-second.

One can also define the *light cone at* $\mathbb{O}$, which we denote by $C_{\mathbb{O},c}$, to be the three-dimensional cone in $T_{\mathbb{O}}(V)$ that is defined by all of the lines through $\mathbb{O}$ that intersect $S_c$. One can also think of a 3-cone in $T_{\mathbb{O}}(V)$ as a 2-sphere in $PT_{\mathbb{O}}(V)$, hence, in $\mathbb{R}P^3$.

Intuitively, when one allows $c$ – hence, the radius of $S_c$ – to become infinite, the effect is for the light sphere $S_c$ to approach a two-dimensional hyperplane at infinity in $\Pi_3$. Similarly, $C_{\mathbb{O},c}$ approaches the hyperplane at infinity in $V$.

In order to make this rigorous, we first look at the equation for the light cone $r = ct$, which is equivalent to:

$$0 = t^2 - \frac{1}{c^2} r^2 , \tag{2.7}$$

for positive $t$.

If we define the proper time interval (arc-length) along a line through $\mathbb{O}$ to the chosen point whose (homogeneous) coordinates are $(t, r)$ as:

$$\tau^2 = t^2 - \frac{1}{c^2} r^2 , \tag{2.8}$$

then we can also define a change of parameterization of this line from $t$ to $\tau$ by the diffeomorphism:

$$\sigma: \mathbb{R} \to \mathbb{R}, \qquad t \mapsto \tau = \sqrt{t^2 - (r/c)^2} . \tag{2.9}$$

We refer to the hyperboloid of two sheets that is defined by $\tau = 1$ as the *indicatrix* of the Minkowski space that is defined by $V$ and $\tau$. Hence, if the vectors of Minkowski space are regarded as the positions of spacetime world-points they will all be reachable from the origin by line segments whose proper time length is one. If the vectors of Minkowski space are velocity vectors then all of the vectors that lie on the indicatrix will have unit speed. We shall call the sheet that intersects the positive $t$ axis the *future* sheet and the other one the *past* sheet.

It becomes immediately clear that in the limit as $c$ becomes indefinitely large the light cone ($\tau = 0$) converges to the $t = 0$ hyperplane and the future sheet of the indicatrix converges to the $t = 1$ hyperplane.

Now let us look at how the point $\mathbf{x} = (c\tau, r)$ projects onto the point $\mathbf{x}' = (ct', r')$ on a (future) $\tau$ = constant hyperboloid by way of a line through the origin. The slope of the line is $r/c\tau$; and one notes that as $c$ goes to infinity the line becomes parallel to the $ct$-axis, which is consistent with projection from a point at infinity. Hence, we must have:

$$\frac{r}{\tau} = \frac{r'}{t'} . \tag{2.10}$$

Since the point $\mathbf{x}'$ is on a $\tau$ = constant hyperboloid, its coordinates must satisfy:

$$\tau^2 = (t')^2 - \frac{1}{c^2} (r')^2 . \tag{2.11}$$

By substitution of (2.10) in (2.11) and solving for $t'$ and then for $r'$, we get:

$$t' = \frac{\tau}{\sqrt{1 - \left(\frac{r}{c\tau}\right)^2}} , \qquad r' = \frac{r}{\sqrt{1 - \left(\frac{r}{c\tau}\right)^2}} . \tag{2.12}$$

This result is also entirely consistent with our having given the line of projection the homogeneous coordinates ($\sqrt{1 - (r/c\tau)^2}$ , $\tau$, $r$), which give the aforementioned values of $t'$ and $r'$ for the corresponding inhomogeneous coordinates.

We illustrate this projection in Fig. 4.

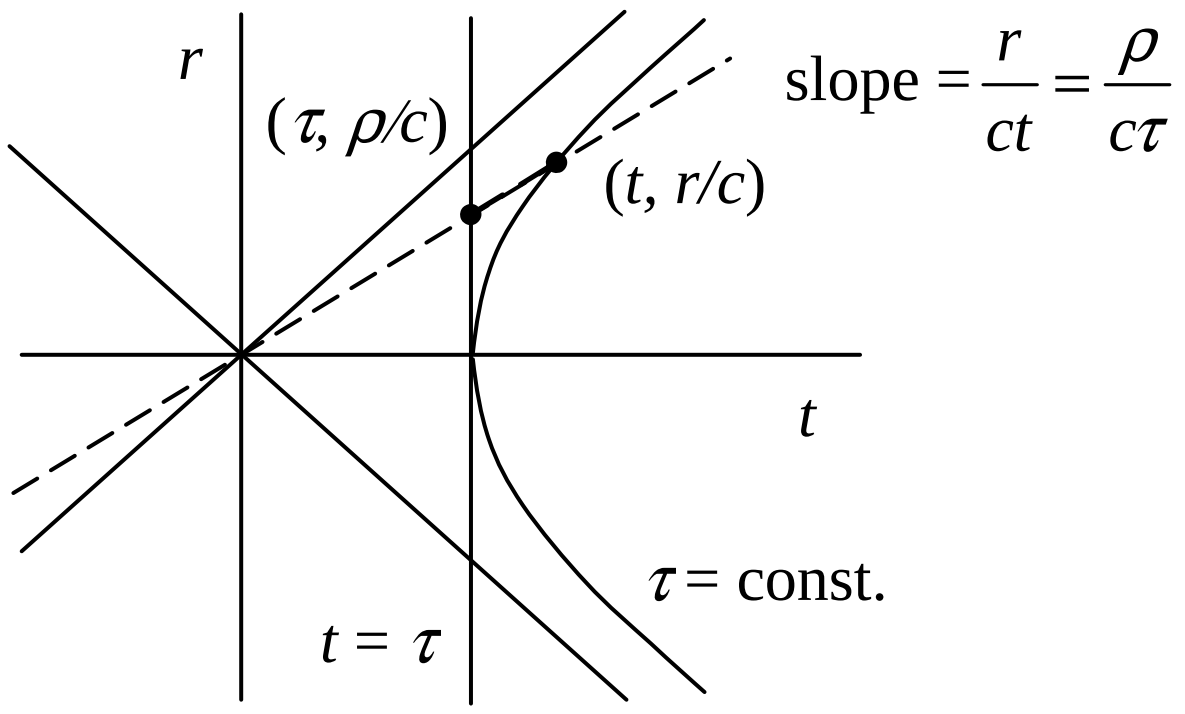

Fig. 4. Projection from a proper time hyperboloid to a time hyperplane.

We shall now see that a similar situation applies to the spatial components of a velocity vector in the rest frame of an observer.

### 2.5 The projection of velocity vectors into rest spaces

The key to making the connection between the last section and conventional special relativity is to note that the process of going from a four-velocity vector **u** that is parameterized by proper time:

$$u^\mu = \frac{dx^\mu}{d\tau}, \qquad (cu^0)^2 - (u^1)^2 - (u^2)^2 - (u^3)^2 = c^2, \tag{2.13}$$

to the spacelike 3-velocity **v** = $dx^i/dt$ that is of interest to experimental physics, one does not simply project orthogonally – i.e., drop the $u^0$ component – but one must also convert to a *different parameterization of the curve*, namely, the parameterization by the time component *t*, where $x^0 = ct$. This means dividing the spatial components of **u** by:

$$\beta = \frac{dt}{d\tau} = \left(1 - \frac{v^2}{c^2}\right)^{-1/2} = u^0. \tag{2.14}$$

The components of the resulting velocity vector have the form of the inhomogeneous coordinates of the line in $\mathbb{R}P^3$ that **u** generates:

$$(u^0, u^i) \mapsto \left(\frac{u^i}{u^0}\right), \tag{2.15}$$

which means that if we set $v^i = u^i/u^0$ then we get:

$$v^i = \left(1 - \frac{v^2}{c^2}\right)^{1/2} u^i, \tag{2.16}$$

which makes physical sense.

### 2.6 1+3 decompositions of spacetime

Since we have made frequent appeals to decompositions of $\mathbb{R}^4$ into essentially a proper time line and a complementary spacelike rest space, it is illuminating to examine the nature the space of all such decompositions.

Physically, a decomposition of spacetime into time and space starts with a choice of a physically-defined motion that one characterizes by its velocity vector field **t**. Note that reality generally implies that the support of **t**, which we denote by $\mathcal{S}(\mathbf{t})$, is not all of $\mathbb{R}^4$, but only a "world-tube" that is foliated into a congruence of world lines. The line field [**t**] that is generated by **t** on $\mathcal{S}(\mathbf{t})$ represents a choice of proper time direction at each point of $\mathcal{S}(\mathbf{t})$. However, in the absence of a metric on the tangent spaces (or at least a norm) one cannot define a preferred proper time parameterization for a given world line since one cannot define unit tangent vectors. Furthermore, whether one can define a parameterization that is common to all the world lines – hence, a "proper time function" on $\mathcal{S}(\mathbf{t})$ – is related to the question of whether the flow of **t** can be defined more than just locally at any point of $\mathcal{S}(\mathbf{t})$.

We now let $\mathbb{R}^4$ represent any of the tangent spaces to $\mathcal{S}(\mathbf{t})$ so that we can stay within the scope of special relativity. Eventually, we intend to apply the intuition thus obtained to the more general case of tangent spaces to four-dimensional manifolds that are not diffeomorphic to vector spaces, but first we wish to establish the physical notions that correspond to the projective geometric ones.

The association of a rest space $\Pi_3$ that makes $\mathbb{R}^4 = [\mathbf{t}] \oplus \Pi_3$ would be straightforward in Minkowski space, since one could use the orthocomplement of [**t**] for $\Pi_3$, unless [**t**] were lightlike. However, in the absence of an orthogonal structure, one can only choose $\Pi_3$ arbitrarily, except for the condition [**t**] cannot lie in $\Pi_3$.

Clearly, a line [**t**] through the origin in $\mathbb{R}^4$ is simply an element of $\mathbb{R}P^3$ and a hyperplane $\Pi_3$ can also be associated with a line through the origin in $\mathbb{R}^{4*}$, which we understand to be the dual space to $\mathbb{R}^4$, i.e., the vector space of linear functionals on $\mathbb{R}^4$. Any element $\alpha \in \mathbb{R}^{4*}$ determines a unique hyperplane in $\mathbb{R}^4$ by way of the space of all vectors $\mathbf{v} \in \mathbb{R}^4$ that are annihilated by $\alpha$: i.e., $\alpha(\mathbf{v}) = 0$. Indeed, the same hyperplane is defined by any other $\lambda\alpha \in \mathbb{R}^{4*}$, where $\lambda$ is a non-zero scalar, which says that really any hyperplane in $\mathbb{R}^4$ is associated with a unique line through the origin of $\mathbb{R}^{4*}$.

We note that although any frame $\mathbf{e}_\mu$, $\mu = 0, 1, 2, 3$ in $\mathbb{R}^4$ defines a unique reciprocal coframe $\theta^\mu$, $\mu = 0, 1, 2, 3$ by the requirement that $\theta^\mu(\mathbf{e}_\nu) = \delta^\mu_\nu$, unless one specifies such a choice of frame, there is no *canonical* isomorphism of $\mathbb{R}^4$ with $\mathbb{R}^{4*}$. When projectivized, this also says that there is no canonical projective isomorphism between $\mathbb{R}P^3$ and $\mathbb{R}P^{3*}$. Otherwise stated, although every line through the origin of $\mathbb{R}^4$ can be complemented by *some* linear hyperplane, such a complement is not generally unique [2].

Thus, the hyperplane $\Pi_3$ can be regarded as an element $[\theta] \in \mathbb{R}P^{3*}$, where $[\theta]$ is the unique line through the origin of $\mathbb{R}^{4*}$ that annihilates $\Pi_3$. Hence, in general, a 1+3 decomposition of $\mathbb{R}^4$ is associated with an element $([\mathbf{t}], [\theta])$ of $\mathbb{R}P^3 \times \mathbb{R}P^{3*}$. However, one must keep in mind that $[\mathbf{t}]$ cannot be contained in the 3-plane $\Pi_3$. Hence, one cannot have $[\theta][\mathbf{t}] = 0$, by which we mean that for any covector $\theta$ and vector $\mathbf{t}$ that generate these lines, we must not have $\theta(\mathbf{t}) = 0$.

---

[2] In the language of lattice theory, this amounts to the statement that the lattice of subspaces of any projective space is complemented, but not orthocomplemented.

We then define the hypersurface $Z$ in $\mathbb{R}P^3 \times \mathbb{R}P^{3*}$ by $Z = \{([\mathbf{t}], [\theta]) \mid [\theta][\mathbf{t}] = 0\}$ and see that the manifold of all 1+3 decompositions of $\mathbb{R}^4$ is simply $\mathbb{R}P^3 \times \mathbb{R}P^{3*} - Z$, which consists of two connected components that have local charts of the form $\mathbb{R}^3 \times \mathbb{R}^3$.

For the sake of completeness, we point out that the hypersurface $Z$ is also the total space of the tautological line bundle over the Grassmanian manifold of 3-planes in $\mathbb{R}^4$, $G_{3,4} = \mathbb{R}P^{3*}$, which consists of all pairs of the form: (3-plane, line in that 3-plane). One could also regard $Z$ as fibered over $\mathbb{R}P^3$ by thinking of it as consisting of all pairs of the form: (line, 3-plane that contains it); the fibers of this bundle then consist of 3-planes.

If one wishes that every line $[\mathbf{t}] \in \mathbb{R}P^3$ be associated with a *unique* hyperplane $[\theta] \in \mathbb{R}P^{3*}$ then one could introduce that as a hypothesis. It is weaker that the existence of an orthogonal structure, since if one assumes that the association of lines in $\mathbb{R}P^3$ with hyperplanes in $\mathbb{R}P^{3*}$ is a one-to-one projective correspondence – so it also takes the 2-planes and 3-planes in $\mathbb{R}P^3$ that are spanned by pairs and triples of lines to 2-planes and 3-planes in $\mathbb{R}P^{3*}$ that are spanned by their images – then one has defined a *correlation* on $\mathbb{R}P^3$. Such a projective isomorphism can be represented in homogeneous coordinates by an equivalence class $[C]$ of scalar multiples of a linear isomorphism $C$: $\mathbb{R}^4 \to \mathbb{R}^{4*}$. We represent the projectivization of $C$ to $[C]$ by the following commutative diagram:

$$\begin{array}{ccc} \mathbb{R}^4 & \xrightarrow{C} & \mathbb{R}^{4*} \\ \downarrow & & \downarrow \\ \mathbb{R}P^3 & \xrightarrow{[C]} & \mathbb{R}P^{3*} \end{array}$$

in which the vertical arrows are the canonical projections.

The isomorphism $C$ also defines a non-degenerate bilinear form on $\mathbb{R}^4$ by way of:

$$C(\mathbf{v}, \mathbf{w}) \equiv C(\mathbf{v})(\mathbf{w}) . \tag{2.17}$$

This projectivizes to a map $[C]$ from $\mathbb{R}P^3 \times \mathbb{R}P^3$ to the set $\mathbb{R}P^0 = \{0, \neq 0\}$of "projective scalars," in which we have added 0 as the "point at infinity," and we abbreviate the expression $[C]([\mathbf{v}], [\mathbf{w}]) = (\neq 0)$ to simply $[C]([\mathbf{v}], [\mathbf{w}]) \neq 0$ . The projective geometric interpretation of the statement $[C]([\mathbf{v}], [\mathbf{w}]) = 0$ is that the line $[\mathbf{w}]$ is incident with the hyperplane $[C]([\mathbf{v}])$.

The bilinear form $C$ – or the map $[C]$, for that matter – does not have to be symmetric, in the general case. Indeed, if it were it would define an orthogonal structure on $\mathbb{R}^4$, hence, a *polarity* on $\mathbb{R}P^3$, to use the language of projective geometry. Of course, $C$ can be polarized:

$$C = C_+ + C_- = \tfrac{1}{2}(C + C^{\mathrm{T}}) + \tfrac{1}{2}(C - C^{\mathrm{T}}) \tag{2.18}$$

into a symmetric part $C_+$ and an anti-symmetric part $C_-$ ; we have introduced the notation $C^{\mathrm{T}}$ to refer to the transposed correlation:

$$C^{\mathrm{T}}(\mathbf{v}, \mathbf{w}) = C(\mathbf{w})(\mathbf{v}). \tag{2.19}$$

One notes that $C$ then defines a quadratic form on $\mathbb{R}^4$ :

$$Q(\mathbf{v}) = C(\mathbf{v}, \mathbf{v}) = C_+(\mathbf{v}, \mathbf{v}) , \tag{2.20}$$

even when it is asymmetric since $C_-(\mathbf{v}, \mathbf{v}) = 0$ in any case. However, unless $C_- = 0$, one has no guarantee that $Q$ is non-degenerate.

Assuming that $Q$ is non-degenerate, we can define the *absolute quadric hypersurface* in $\mathbb{R}^4$ relative to $C$ by way of the set of all *isotropic* vectors, for which $Q(\mathbf{v}) = 0$. Under projectivization, one obtains the absolute quadric surface of isotropic lines in $\mathbb{R}\mathrm{P}^3$. Such a line will then be incident with its correlated hyperplane under $[C]$.

One then sees that if one defines a correlated hyperplane to $[\mathbf{t}]$ by way of $[\theta] = [C]([\mathbf{t}])$ then the statement that $[\mathbf{t}]$ is not incident with the hyperplane $[\theta]$ is equivalent to the statement that $[\theta]([\mathbf{t}]) \neq 0$, which is equivalent to $Q([\mathbf{v}]) \neq 0$. Hence, we obtain the consequence that *no isotropic line can define a 1+3 decomposition of* $\mathbb{R}^4$, which is a generalization of the fact that light rays do not define rest frames or proper time parameterizations.

If one defines a correlation on $\mathbb{R}^4$ whose symmetric part defines the Minkowski scalar product then clearly the only way that more general correlations would have physical significance is if there were a physical context in which the anti-symmetric part played a key role. We shall see later that such is the case when one defines a correlation on the space of bivectors, which will also be given a complex structure that follows naturally from the structure of electromagnetism.

Now, the only manner in which a choice of 1+3 decomposition of $\mathbb{R}^4$ is physically acceptable is if all such choices are equivalent in some sense, or at least, there is a well-defined equivalence class of non-unique choices. This brings us to the issue of groups of transformations that act on 1+3 decompositions, so we now discuss that matter in more detail.

## 3 Projective transformations

So far, we have identified three key choices in the process of setting up the projective geometric machinery of spacetime: a choice of rest space, a choice of proper time line, and a choice of parameterization for that line. The natural question to ask is then: "to what extent can we define geometric objects on spacetime that are independent of these choices?" Hence, we need to establish groups of transformations that are defined by making other choices. Furthermore, we need to establish the corresponding physical significance of these transformations. Finally, we need to relate the more general picture of projective geometry to the well-established picture of special relativity.

Now, the first two objects – viz., a choice of rest space and proper time line – can be combined into a choice of projective frame for $V$. Since this involves a choice of four (in this case) linearly independent lines, one sees that it also involves a choice of four independent choices of parameterization for the lines. Hence, we need to treat both of the groups thus defined, as well as the extent to which they relate to each other. For instance, it is clear that any reparametrization of all four lines of the projective frame by four constant scalar factors is equivalent to a projective transformation of shear type.

### 3.1 Transformations between projective frames

A *projective frame* ([3]) for the projective space $\mathbb{R}P^n$ is a set of $n+1$ points $[\mathbf{e}_\mu] \in \mathbb{R}P^n$ that "span" all of $\mathbb{R}P^n$. The easiest way to clarify this definition is by looking at any set of $n+1$ vectors $\mathbf{e}_\mu \in \mathbb{R}^{n+1}$ that project onto $[\mathbf{e}_\mu]$. One then says that the $[\mathbf{e}_\mu]$ span $\mathbb{R}P^n$ iff the $\mathbf{e}_\mu$ span $\mathbb{R}^{n+1}$. Since any other choice of representative $n+1$-frame in $\mathbb{R}^{n+1}$ for the $[\mathbf{e}_\mu]$ is related to the first one by an invertible transformation, this property is independent of the choice of representative $n+1$-frame.

For the sake of working with matrix representations of the groups involved, it is more convenient to work with the n+1-frame $\mathbf{e}_\mu$ that relates to the homogeneous coordinates of any line in $\mathbb{R}P^n$.

There are four elementary transformations of $GL(n+1; \mathbb{R})$ that can affect the frame $\mathbf{e}_\mu$:

*i*) *Homotheties:* Such a transformation takes $\mathbf{e}_0$ to $\lambda\mathbf{e}_0$ and leaves the other members alone.

*ii*) *Translations:* This time $\mathbf{e}_0$ stays fixed, but $\mathbf{e}_i$ goes to $\mathbf{e}_i + \mathbf{a}$ .

*iii*) *Inversions:* Such transformations take $\mathbf{e}_0$ to $\mathbf{e}_0 + \mathbf{b}$, and leave the other frame members fixed.

*iv) Linear transformations:* These transformations leave $\mathbf{e}_0$ fixed and take $\mathbf{e}_i$ to $A^i_j\,\mathbf{e}_i$.

We can assemble these four types of transformations into a partitioned $(n+1)\times(n+1)$ real invertible matrix:

$$A = \left[\begin{array}{c|c} \lambda & b_j \\ \hline a^i & A^i_j \end{array}\right]. \tag{3.1}$$

Such a matrix $A$ acts on the homogeneous coordinates $(x^0, \ldots, x^n)$ of a point $[x] \in \mathbb{R}P^n$ by matrix multiplication, although one must then note that all non-zero scalar multiples of $A^\mu_\nu x^\nu$ will produce the same $[x]$ when projected onto $\mathbb{R}P^n$. Hence, one must regard the entire equivalence class of $[A]$ – i.e., the intersection of the line through the origin of $\mathbb{R}^{(n+1)^2}$ that $A$ defines with $GL(n+1; \mathbb{R})$ – as constituting the same transformation of $\mathbb{R}P^n$. One calls the quotient of $GL(n+1; \mathbb{R})$ by this equivalence the *projective linear group*

[3] For the approach to projective differential geometry using moving frames, see Cartan [**10**], or, more recently, Akivis and Goldberg [**11**]. A classic reference that took the Lie approach of invariants of symmetries of systems of differential equations is Wilczynski [**12**].

$PGL(n;\mathbb{R})$ in $n$ (projective) dimensions. Since every equivalence class $[A]$ contains a unique element $\tilde{A}$ with unit determinant, namely:

$$\tilde{A} = \left(\frac{1}{\det(A)}\right)^{n+1} A\,, \tag{3.2}$$

one can also regard $PGL(n;\mathbb{R})$ as isomorphic to $SL(n+1;\mathbb{R})$.

By projection, the group $SL(n+1;\mathbb{R})$ acts nonlinearly on the inhomogeneous coordinates of a point in $\mathbb{R}\mathrm{P}^n$ by way of *linear fractional transformations:*

$$AX^i = \frac{a^i + A^i_j X^j}{\lambda + b_j X^j}\,. \tag{3.3}$$

Note, in particular, that a pure dilatation $\lambda$ now acts on $X^i$ as $1/\lambda$.

An inversion $X^i \mapsto X^i / (1 + b_i X^i)$ has all of the points of the linear hyperplane $b_i X^i = 0$ for its fixed points. Hence, if we have a line $L$ in $\mathbb{R}^n$ that intersects this hyperplane at a point then the a point of $L$ that is at a distance $r$ on one side the hyperplane will get mapped to a point of $L$ that is at a distance of $1/r$ on the other side of it. One can also characterize the denominator in (3.3) as representing an inversion through the affine hyperplane $\lambda + b_jX^j = 1$.

We also see that if we choose to regard a projective frame as a set of $n+1$ points in $\mathbb{R}\mathrm{P}^n$ then the action of the projective linear group on such a frame is rather complicated compared to the linear action of $SL(n+1;\mathbb{R})$ on any $(n+1)$-frame that projects onto it.

We now restrict our scope to $n = 3$.

In order to relate the action of projective transformations on a homogeneous 4-frame in $\mathbb{R}^4$ to transformations of the affine 3-plane $\Pi_3$ and proper time line $L$, we assume that $L$ is generated by the frame vector $\mathbf{e}_0$ and that $\Pi_3$ is the translate by $\mathbf{e}_0$ of the linear 3-plane that is spanned by the remaining $\mathbf{e}_i$, namely, the hyperplane at infinity, $\Pi_\infty$.

In this context, we can describe the projective transformation in terms of their effect on $L$ and $\Pi_3$: Homotheties take the vectors of $L$ to other vectors of $L$ and the vectors of $\Pi_3$ to other vectors of $\Pi_3$. Translations take vectors of $\Pi_3$ to other vectors of $\Pi_3$ while leaving $L$ untouched. Conversely, inversions change the direction of $L$ while leaving the vectors of $\Pi_3$ untouched. Linear transformations take vectors of $\Pi_3$ to other vectors of $\Pi_3$ while fixing both the origin of $\Pi_3$ and the line $L$.

If one looks at the subgroup of $GL(4;\mathbb{R})$ that takes a vector of $\Pi_\infty$ to another vector of $\Pi_\infty$ then an element of this subgroup must take any point with the homogeneous coordinate $x^0 = 0$ to another point with $x^0 = 0$, which says that $0 = b_i x^i$ for any $(x^1, x^2, x^3)$, which can only be satisfied if $b_i = 0$. This says that all of the four elementary transformations except for the inversions preserve the hyperplane at infinity. This subgroup is then the affine group of $\Pi_3$ plus the homotheties. Hence, the role of inversions also amounts to a change in the choice of hyperplane at infinity ([4]), or, as we saw, a change in the choice of *L*. Later, we shall see that inversions are also related to boost transformations in Minkowski space – i.e., transformations to frames with constant relative velocity, - as well as transformation to constantly accelerated frames.

If we look at the subgroup of $GL(4; \mathbb{R})$ that takes a vector of $\Pi_3$ to another vector of $\Pi_3$ then an element of this subgroup must take any point with the homogeneous coordinate $x^0 = 1$ to another point with $x^0 = 1$, which says that $1 = \lambda + b_i x^i$ for any $(x^1, x^2, x^3)$, which implies that $b_i = 0$ and $\lambda = 1$. The subgroup thus defined is then the affine group of $\Pi_3$ .

### 3.2 Transformations of parameterizations

Since a parameterization for a line $L \in PV$ is a diffeomorphism $\sigma: L \to \mathbb{R}$, any other parameterization $\sigma'$ of *L* will be obtained by composing $\sigma$ with a unique diffeomorphism of $\mathbb{R}$ to itself, namely, $\sigma' \cdot \sigma^{-1}$. Hence, when one chooses an initial parameterization for a given line there is a one-to-one correspondence between the other parameterizations of that line and the elements of the group Diff($\mathbb{R}$), which we now discuss.

Since the derivative $dy/dx$ of any diffeomorphism $y = y(x)$ of $\mathbb{R}$ to itself cannot vanish for any *x*, that derivative can either be everywhere positive or everywhere negative on $\mathbb{R}$. Hence, Diff($\mathbb{R}$) has two components: the subset $\text{Diff}_+(\mathbb{R})$, which contains the identity transformation, and which consists of non-decreasing functions on $\mathbb{R}$, and the subset $\text{Diff}_-(\mathbb{R})$, which consists of non-increasing functions and is in one-to-one correspondence with $\text{Diff}_+(\mathbb{R})$ by way of the parity inversion of $\mathbb{R}$ that takes every *x* to $-x$. Neither subset is a subgroup, since the inverse of a non-decreasing function is a non-increasing one, and vice versa.

By differentiation, non-decreasing functions are associated with positive functions on $\mathbb{R}$ and non-increasing functions with negative functions, which serve as essentially the infinitesimal generators of the diffeomorphisms. One sees that the uniqueness of this association is only up to an additive integration constant, which must be non-negative or non-positive, respectively. Hence, one can regard an element $\sigma \in \text{Diff}_+(\mathbb{R})$ as being uniquely associated with a pair $(\sigma_0, \ln \lambda) \in \mathbb{R}^+ \times C^\infty_+(\mathbb{R})$, where:

$$\sigma(x) = \sigma_0 + \lambda(x)\, x, \tag{3.4}$$

for every $x \in \mathbb{R}$. One simply sets:

$$\sigma_0 = \sigma(0), \qquad \lambda(x) = \frac{\sigma(x) - \sigma_0}{x}, \qquad (x \neq 0). \tag{3.5}$$

[4] In the early attempts at using projective geometry to account for the Kaluza-Klein model (cf., e.g., Veblen [**13**], and, more recently, Schmutzer [**14**]), the gauge transformations of electromagnetism were thus interpreted as changes of the choice of hyperplane at infinity.

and defines $\lambda(0)$ by continuity; of course, the function $\lambda$ itself must have the property that $\lambda \geq 1$.

Similarly, an element of Diff$_-(\mathbb{R})$ is associated with an element of $\mathbb{R}^- \times C^\infty_-(\mathbb{R})$, and by taking the union Diff$_+(\mathbb{R}) \cup$ Diff$_-(\mathbb{R})$ one sees that any element of Diff$(\mathbb{R})$ is associated with a unique element of $\mathbb{R} \times C^\infty(\mathbb{R})$, although the association is not onto, since we are not using the elements of $\mathbb{R}^+ \times C^\infty_-(\mathbb{R})$ or $\mathbb{R}^- \times C^\infty_+(\mathbb{R})$. The elements of $\mathbb{R} \times C^\infty(\mathbb{R})$ can also be regarded as elements of $C^\infty(\mathbb{R}, \mathbb{R}^2)$ of a particular sort, namely, functions from $\mathbb{R}$ to $\mathbb{R}^2$ whose first component is a constant function.

The group structure of Diff$(\mathbb{R})$ under composition of functions translates into the semi-direct product of the additive group of $\mathbb{R}$ with the multiplicative group of non-negative smooth functions, since if $\sigma(x) = \sigma_0 + \lambda(x)\, x$ and $\sigma'(x) = \sigma'_0 + \lambda'(x) x$ then:

$$(\sigma' \circ \sigma)(x) = [\sigma'_0 + \lambda'(x)\sigma_0] + (\lambda'\lambda)(x)\,. \tag{3.6}$$

One immediately confirms that this multiplication law is non-commutative, since:

$$(\sigma \circ \sigma')(x) = [\sigma_0 + \lambda(x)\sigma'_0] + (\lambda\lambda')(x)\,. \tag{3.7}$$

One can also see the one-to-one correspondence between the elements of Diff$_+(\mathbb{R})$ and non-negative functions simply by noting that any diffeomorphism defines an associated *displacement* function by way of:

$$d(x) \equiv \sigma(x) - x\,. \tag{3.8}$$

As long as $\sigma(x) \geq x$, one will have that $d(x) \geq 0$.

A common finite-dimensional subgroup of Diff$(\mathbb{R})$ that one encounters in applications is the two-dimensional subgroup of *affine reparameterizations*, which take the form of (3.4), only with constant $\lambda$.

If one wishes to simultaneously reparameterize the members of a projective $n+1$-frame independently then one defines an element of the group Diff$(\mathbb{R})^{n+1}$, which is the direct product of $n+1$ copies of Diff$(\mathbb{R})$. Hence, an infinitesimal generator of such a transformation is an element of $\mathbb{R}^{n+1} \times C^\infty(\mathbb{R})^{\,n+1}$, or a particular element of $C^\infty(\mathbb{R}^{\,n+1}, \mathbb{R}^{n+2})$, as above.

### 3.3 Relationships between the two groups

In the previous subsection, it was becoming clear that what we were constructing were subgroups of the group $C^\infty(\mathbb{R}^{n+1}, GL(n+1; \mathbb{R}))$, which acts on $\mathbb{R}^{n+1}$ by matrix multiplication when one regards an element of the group as an invertible $n+1$ by $n+1$ matrix of real functions on $\mathbb{R}^{n+1}$; this is also the form that the differential $D\phi$ of a coordinate transition function $\phi: \mathbb{R}^{n+1} \to \mathbb{R}^{n+1}$ is presumed to take. When $\phi$ is an invertible linear transformation, one has that $D\phi$ agrees with $\phi$. One has that the group $C^\infty(\mathbb{R}^{n+1}, GL(n+1; \mathbb{R}))$ contains $GL(n+1; \mathbb{R})$ as the subgroup of all constant maps from $\mathbb{R}^{n+1}$ to $GL(n+1; \mathbb{R})$. Hence, we conclude that the scope of the group of re-parameterizations of all lines in the affine space $\mathbb{A}^{n+1}$ is considerably larger than that of mere projective transformations, which are included as a finite-dimensional subgroup.

In order to actually reconstruct elements of $GL(n+1; \mathbb{R})$ by simply reparameterizing the lines in the affine space $\mathbb{A}^{n+1}$ there is a subtlety that we must address: these lines do not come with a canonical parameterization, but only the arc-length parameterizations that they implicitly inherit from a metric, and, even then, the choice of origin is still arbitrary. That is, if $d$: $\mathbb{A}^{n+1} \times \mathbb{A}^{n+1} \to \mathbb{R}$ is a metric on $\mathbb{A}^{n+1}$ and $l$ is an oriented line in $\mathbb{A}^{n+1}$ then if one chooses a point $P$ on $l$ that will serve as origin then one can define a parameterization of $l$ by way of $s(x) \equiv \pm\, d(P, x)$ , in which the sign is determined by the orientation of $l$. One is further assuming various types of consistency and homogeneity of the parameterizations of lines through different points of $\mathbb{A}^{n+1}$ that project from the same set of points in $\mathbb{A}^{n+1}$.

If we wish to make our definitions independently of a choice of metric on $T(\mathbb{A}^{n+1})$ then we need to abstract the process of assigning a parameterization to each line in $\mathbb{A}^{n+1}$ .

The set of all lines in $\mathbb{A}^{n+1}$ is really the projectivized tangent bundle to $\mathbb{A}^{n+1}$, viz., $PT(\mathbb{A}^{n+1})$, which also defines a tautological (real) line bundle $L(PT) \to PT(\mathbb{A}^{n+1})$ by way of the projection $T(\mathbb{A}^{n+1}) \to PT(\mathbb{A}^{n+1})$. That is, the fiber $L_l(PT)$ over any line $l_x$ through a point $x \in \mathbb{A}^{n+1}$ is the set of points in $\mathbb{A}^{n+1}$ that make up the line $l_x$ . Note that although the total space $T(\mathbb{A}^{n+1})$ of this fibration is diffeomorphic to $\mathbb{A}^{n+1} \times \mathbb{R}^{n+1}$, there is still a different projection at each point of $\mathbb{A}^{n+1}$ , so the fibration is not merely projection onto the first factor of the Cartesian product.

Hence, we need to associate each line $l_x$ in $PT(\mathbb{A}^{n+1})$ with a diffeomorphism of the fiber $L_l(PT)$ with $\mathbb{R}$. This means that we need to define the bundle $\Pi\mathrm{L}(PT) \to PT(\mathbb{A}^{n+1})$ whose fibers are diffeomorphisms of the fibers of $L(PT)$ with $\mathbb{R}$. A choice of global parameterization is then a global section of this bundle. Furthermore, any two choices of section would be related by a map from $PT(\mathbb{A}^{n+1})$ to $\mathrm{Diff}(\mathbb{R})$.

Of course, all of this sounds suspiciously reminiscent of defining a global "gauge" on $PT(\mathbb{A}^{n+1})$, but keep in mind that despite the one-dimensionality of the fibers of $L(PT)$ our "gauge group" is really the *infinite*-dimensional group $\mathrm{Diff}(\mathbb{R})$. Hence, a local gauge – or "parameterization frame" – has an infinitude of "legs." Furthermore, since $PT(\mathbb{A}^{n+1})$ is diffeomorphic to $\mathbb{R}^{n+1} \times \mathbb{R}\mathrm{P}^n$, which is homotopically equivalent to $\mathbb{R}\mathrm{P}^n$, which is not even simply connected, much less contractible, one might have to accept that the fibration $\Pi\mathrm{L}(PT) \to PT(\mathbb{A}^{n+1})$ is not trivializable, so global gauges might not exist ([5]), even for the elementary case in which we started out with an affine space.

A further condition that one might impose on a choice of parameterization is one of *consistency*. That is, since the same set of points in $\mathbb{A}^{n+1}$ project onto a different line through each of those points, one might demand that the same parameterization of that point set in $\mathbb{A}^{n+1}$ apply to each of the lines in $PT(\mathbb{A}^{n+1})$ that it projects onto. One might also impose the condition of *homogeneity*, i.e., the condition that the parameterizations are consistent with the action of the translation group $\mathbb{R}^{n+1}$ on $\mathbb{A}^{n+1}$ ([6]). That is, when all of the points of $\mathbb{A}^{n+1}$ are translated by the same vector $\mathbf{v} \in \mathbb{R}^{n+1}$ the effect on all lines in $PT(\mathbb{A}^{n+1})$ that are invariant under the action of $\mathbf{v}$ is a common translation of the parameter values in $\mathbb{R}$. Indeed, this restriction is clearly stronger than mere consistency, since one could use the action of translations along any given line to force consistency of the various parameterizations of the lines it projects to in $PT(\mathbb{A}^{n+1})$.

[5] In obstruction theory, the key notion would be the first non-vanishing homotopy group of $\mathrm{Diff}(\mathbb{R})$, but we shall not pursue this direction, for now.

One could further restrict oneself to homogeneity under the action of the some larger group, such as the Galilei group or Poincaré group, depending upon what further structures one has imposed on $T(\mathbb{A}^{n+1})$. For instance, one might wish that the parameterizations of all lines through a given point are invariant under all rotations, if one has defined a Riemannian structure. Since the rotation group $SO(n+1; \mathbb{R})$ acts transitively on $\mathbb{R}P^n$ it would then be sufficient to define a single parameterization on some chosen line in some chosen fiber of $PT(\mathbb{A}^{n+1})$ in order to define the parameterizations of all the other lines in $PT(\mathbb{A}^{n+1})$.

With the restriction of translational homogeneity, one essentially collapses the bundle $\Pi L(PT) \to PT(\mathbb{A}^{n+1})$ to the bundle $\Pi L_P(PT) \to \mathbb{R}P^n = PT_P(\mathbb{A}^{n+1})$, at least when one chooses a reference point $P$ to serve as "origin." (This the essence of the process of collapsing the full group $C^\infty(\mathbb{R}^{n+1}, GL(n+1; \mathbb{R}))$ to the $GL(n+1; \mathbb{R})$ subgroup of constant maps.)

Eventually, we intend to apply the methods that are discussed in this article in the context of affine spaces and their projectivizations to the more modern general relativistically mandated context of differentiable manifolds. The methods discussed herein would then, presumably, apply to the projectivized tangent spaces the spacetime manifold. One could then regard the affine space $\mathbb{A}^{n+1}$ as a special case in which the manifold has enough symmetry to be a homogeneous space under the action of the translation group, or simply the group manifold for the translation group, when one chooses a point of $\mathbb{A}^{n+1}$. However, one of the recurring problems of general relativistic physics is that although linear transformations of the tangent spaces to the spacetime manifold are well-defined, nevertheless, there are fundamental differences between the displacements of *points* in $\mathbb{A}^{n+1}$ by vectors in $\mathbb{R}^{n+1}$ and the parallel displacements of *tangent vectors* to a more general manifold. In particular, one must define displacements along curves, and although any two points of $\mathbb{A}^{n+1}$ can be connected by a unique curve – viz., the line segment between them – the same is not always true of manifolds that are given curved connections. Hence, we humbly defer the application of the present discussion to more general manifolds to a later work, and try to gain as much intuitive sense of what is going on in the more tractable case of affine spaces and their projective spaces.

### 3.4 The reduction to the Lorentz group

We described the transition from $\mathbb{R}^4$ to Minkowski space above as a rescaling of the proper time parameter that depended on the points of $\mathbb{R}^4$ that the proper time line passed through. As a consequence, the hyperplane at infinity of $\mathbb{R}\mathrm{P}^3$ – viz., $x^0 = 0$ – was deformed into the light cone of Minkowski space, or, equivalently, the unit sphere in the affine hyperplane $x^0 = 1$, and the latter affine hyperplane was deformed into the indicatrix of Minkowski space; i.e., the $\tau = 1$ hyperboloid. The effect on the lines through the origin was to transform them into other lines through the origin that lived in the interior region bounded by the light cone ([7]).

[6] Of course, anyone who is thinking ahead to the generalization of projective differential geometry is skeptical of any assumption that the manifold one is starting with has any such homogeneity.

Since we could define subgroups of $GL(4; \mathbb{R})$ by looking at transformations that preserved the hyperplane at infinity and the affine hyperplane $x^0 = 1$, it is natural to inquire what the corresponding subgroups of $GL(4; \mathbb{R})$ are that preserve the light cone and indicatrix.

Preserving the light cone in Minkowski space is equivalent to preserving the unit sphere in $\mathbb{R}\mathrm{P}^3$. Since $GL(4; \mathbb{R})$ acts on $\mathbb{R}\mathrm{P}^3$ by way of linear fractional transformations on the inhomogeneous coordinates, we can formulate the problem as one of looking for all $A \in GL(4; \mathbb{R})$ such that if:

$$1 = g_{ij}X^iX^j \tag{3.9}$$

then:

$$1 = g_{ij}\frac{a^i + A^i_k X^k}{\lambda + b_k X^k}\frac{a^j + A^j_l X^l}{\lambda + b_l X^l} \tag{3.10}$$

or:

$$(\lambda + b_kX^k)^2 = g_{ij}(a^ia^j + 2a^iA^j_kX^k + A^i_kA^j_lX^kX^l). \tag{3.11}$$

The matrix components are not independent, since if $a^i = 0$ then this reduces to:

$$(\lambda + b_kX^k)^2 = g_{ij}A^i_kA^j_lX^kX^l, \tag{3.12}$$

which can only be satisfied for all $X^i$ that satisfy (3.9) if $b_i = 0$, as well. The remaining condition on $A$ says that it must take the form of the non-zero scalar $\pm\lambda$ times a unique Euclidian rotation matrix. Hence, the subgroup that is defined by $a^i = 0$ is the direct product of the multiplicative group $\mathbb{R}^*$ with $O(3; \mathbb{R})$. Conversely, if $b_i = 0$ then one must have $a^i = 0$, as well.

Consequently, we see that a non-zero translation in the affine hyperplane $\Pi_3$ ($x^0 = 1$) must be compensated for by a non-zero inversion through the affine hyperplane defined by $\lambda$ and $b_i$. This is exactly what the boosts do.

The elementary boosts in the $x$, $y$, and $z$ directions, namely:

$$B_x = \begin{bmatrix} \cosh\alpha & \sinh\alpha & 0 & 0 \\ \sinh\alpha & \cosh\alpha & 0 & 0 \\ 0 & 0 & 1 & 0 \\ 0 & 0 & 0 & 1 \end{bmatrix}, \quad B_y = \begin{bmatrix} \cosh\alpha & 0 & \sinh\alpha & 0 \\ 0 & 1 & 0 & 0 \\ \sinh\alpha & 0 & \cosh\alpha & 0 \\ 0 & 0 & 0 & 1 \end{bmatrix},$$

[7] That is, the disconnected region.

$$B_z = \begin{bmatrix} \cosh\alpha & 0 & 0 & \sinh\alpha \\ 0 & 1 & 0 & 0 \\ 0 & 0 & 1 & 0 \\ \sinh\alpha & 0 & 0 & \cosh\alpha \end{bmatrix}, \tag{3.13}$$

correspond to the linear fractional transformations:

$$B_x X^i = \left( \frac{\sinh\alpha + (\cosh\alpha)X^1}{\cosh\alpha + (\sinh\alpha)X^1}, \frac{X^2}{\cosh\alpha + (\sinh\alpha)X^1}, \frac{X^3}{\cosh\alpha + (\sinh\alpha)X^1} \right), \tag{3.14a}$$

$$B_y X^i = \left( \frac{X^1}{\cosh\alpha + (\sinh\alpha)X^2}, \frac{\sinh\alpha + (\cosh\alpha)X^2}{\cosh\alpha + (\sinh\alpha)X^2}, \frac{X^3}{\cosh\alpha + (\sinh\alpha)X^2} \right), \tag{3.14b}$$

$$B_z X^i = \left( \frac{X^1}{\cosh\alpha + (\sinh\alpha)X^3}, \frac{X^2}{\cosh\alpha + (\sinh\alpha)X^3}, \frac{\sinh\alpha + (\cosh\alpha)X^3}{\cosh\alpha + (\sinh\alpha)X^3} \right), \tag{3.14c}$$

resp.

From these formulas, one can identify:

$$\lambda^2 = \cosh^2\alpha = 1 + a_i a^i, \quad a^i = b_i, \quad (b_i)^2 = A^i_m A^i_m - 1 = \sinh^2\alpha, \quad m = 1, 2, 3, \tag{3.15}$$

in which $A^i_m$ is the 3×3 diagonal matrix that differs from the identity matrix by the replacement of 1 in the $A^m_m$ element by cosh $\alpha$. Direct substitution verifies that this choice does indeed satisfy (3.10). Hence, the boost transformations also preserve the unit sphere in $\Pi_3$ when they act as linear fractional transformations.

Therefore, we have deduced the familiar notion that the subgroup of $GL(4; \mathbb{R})$ that preserves the light cone in Minkowski space is the linear conformal group of the Lorentz group. However, this is not the *full* conformal Lorentz group, which is a subgroup of Diff($\mathbb{R}^4$) that includes nonlinear transformations, such as translations inversions, as well, and one must look for a higher-dimensional space in which to find a faithful linear representation. We shall return to this in the next subsection, but first, we discuss the reduction from the linear conformal Lorentz group to the Lorentz group itself.

Just as the reduction from the direct product of $\mathbb{R}^*$ with the three-dimensional affine group $A(3; \mathbb{R})$ to the $A(3; \mathbb{R})$ itself involves looking at linear transformations of $\mathbb{R}^4$ that preserve the affine hyperplane $\Pi_3$, when we deform the hyperplane at infinity into the light cone and the affine hyperplane $\Pi_3$ into the indicatrix $\mathcal{I}$ ($\tau = 1$), the reduction from the linear conformal Lorentz group to the Lorentz group entails only looking at transformations that preserve $\mathcal{I}$.

This time, it is more convenient to use the homogeneous coordinates, so we are looking for all $A \in GL(4; \mathbb{R})$ such that if $g_{\mu\nu} X^\mu X^\nu = 1$ then:

$$1 = g_{\mu\nu} A^\mu_\sigma A^\nu_\tau X^\sigma X^\tau . \tag{3.16}$$

Clearly, this gives us the Lorentz group, but without a scalar multiple, this time.

We previously pointed out that the projection from a Lorentz-invariant 4-velocity to a spatial 3-velocity in a rest space was actually the projection of homogeneous coordinates to inhomogeneous coordinates. Now, let us look at the effect of applying Lorentz transformations to velocity vectors when they are then projected.

Let **u** be a velocity vector whose inhomogeneous – i.e., spatial – coordinates $v^i$ are 0, so its homogeneous components are $u^\mu = (1, 0, 0, 0)$. The velocity vector that results from a boost transformation of **u** by $A \in SO(3,1)$ has the homogeneous components:

$$\bar{u}^\mu = (A_0^0, A_0^1, A_0^2, A_0^3) \tag{3.17}$$

and the inhomogeneous coordinates:

$$\bar{v}^i = (A_0^1 / A_0^0, A_0^2 / A_0^0, A_0^3 / A_0^0)\,. \tag{3.18}$$

For instance, if the Lorentz transformation is an elementary boost in the $x^1$-direction with a rapidity parameter $\alpha$:

$$B_x = \begin{bmatrix} \cosh\alpha & \sinh\alpha & 0 & 0 \\ \sinh\alpha & \cosh\alpha & 0 & 0 \\ 0 & 0 & 1 & 0 \\ 0 & 0 & 0 & 1 \end{bmatrix}, \qquad \sinh\alpha = v/c, \quad \cosh\alpha = \beta, \tag{3.17}$$

then the resulting spatial velocity has inhomogeneous coordinates:

$$v^i = (v, 0, 0). \tag{3.18}$$

Hence, we see that a boost in that direction really does have the effect of translating the velocity vector in the tangent space, as long as one projects onto the spatial plane from a point that is not at infinity.

Now, let us apply this boost to a velocity vector whose spatial components are $(w, 0, 0)$, so we choose homogeneous components of, say, $(u^0, \sqrt{1-u_0^2}\,, 0, 0)$, which dictates that $w = \sqrt{1-u_0^2}\,/u^0$. If we set $u^0 = \beta' = (1 + w)^{-1/2}$, so $\sqrt{1-u_0^2} = \beta' w$, then the resulting homogeneous components are:

$$\begin{aligned}(\beta' \cosh\alpha + \beta' w \sinh\alpha,\ \beta' \sinh\alpha + \beta' w \cosh\alpha, 0, 0) \\ = (\beta\,\beta' + \beta' wv, \beta v + \beta\,\beta' w\,, 0, 0),\end{aligned} \tag{3.19}$$

and the inhomogeneous ones are:

$$\left(\frac{w + v/\beta}{1 + wv/\beta}, 0, 0\right), \tag{3.20}$$

which is consistent with the usual relativistic “addition of velocities” formula.

Now, if we had applied the elementary boost to the lightlike vector (1, 1, 0, 0) then the result would be (cosh $\alpha$ + sinh $\alpha$, sinh $\alpha$ + cosh $\alpha$, 0, 0), which projects to (1, 0, 0), regardless of the value of $\alpha$. This is consistent with the common knowledge that one cannot accelerate a lightlike velocity vector.

### 3.5 Conformal Lorentz transformations

One can think of a choice of light cone on $\mathbb{R}^4$ – or absolute quadric on $\mathbb{R}P^3$ – as a *conformal* structure on $\mathbb{R}^4$. This defines a subgroup $CO(3, 1)$ of the group Diff($\mathbb{R}^4$) that consists of all diffeomorphisms of $\mathbb{R}^4$ that preserve the conformal structure, and that one calls the *conformal Lorentz group.* Hence, such transformations must all take lightlike tangent lines through the origin of the tangent space at one point to other lightlike tangent lines through the origin of the tangent space at some other point. Clearly, that includes all Lorentz transformations, as well as all homotheties. The product of these groups is a subgroup $CO_0(3, 1)$ of $GL(4; \mathbb{R})$ that we will call the *linear conformal Lorentz group*, but it does not exhaust all of the possibilities. In addition to the aforementioned linear transformations of $\mathbb{R}^4$, there are two other types of nonlinear diffeomorphisms that preserve the light cone: translations and inversions through the indicatrix.

The translations appear because they only affect the choice of origin for a tangent space without affecting the tangent vectors themselves. They define a sort “identity map” – up to parallelism – between tangent spaces.

The inversions through the indicatrix are less obvious. They take the form:

$$\mathbf{x} \mapsto \frac{1}{\eta(\mathbf{x},\mathbf{x})}\mathbf{x} \tag{3.21}$$

for any **x** that is not on the light cone, in which we have represented the Minkowski scalar product by $\eta$ = diag(+1, −1, −1, −1). Moreover, just as one compactifies $\mathbb{R}^4$ with a point at infinity in Euclidian conformal geometry so that the inversion of the origin can be defined, similarly, one can conformally compactify Minkowski space by the addition of a “light cone at infinity” so that one can also define the inversion of points on the light cone to be the corresponding points on the light cone at infinity, and vice versa.

One will note that a point of $\mathbb{R}^4$ is fixed by the inversion (3.21) iff it lies on the indicatrix. If the speed of light were infinite then we would be dealing with the inversions that fix the affine hyperplane whose vectors have homogeneous coordinates with $x^0 = 1$. Furthermore, the spacelike vectors will reverse their directions, as well as change their length to one over their former length.

The translations and inversions have a very deep physical significance, since Engstrom and Zorn [**15**] once showed that the transformation from a frame in Minkowski space to another one such that there is a relative constant acceleration of the two frame origins is a composition of a translation, inversion, and a further translation. Hence, one must consider the ramifications of conformal geometry in order to understand what happens when the relativity of constant velocities is extended to the relativity of constant accelerations.

Although the aforementioned linear conformal Lorentz group can be represented as a subgroup of $GL(4; \mathbb{R})$, the full conformal Lorentz group cannot. In fact, although one can represent the Poincaré group, which is the semi-direct product of the Lorentz group with the four-dimensional translation group, as well as the product of this with the homotheties, in $GL(5; \mathbb{R})$, in a manner that analogous to the previous discussion of projective transformations, nevertheless, the problem arises when one notices that if one represents inversions through the indicatrix in the analogous way then one defines matrices that are functions of the point in $\mathbb{R}^4$ that they are applied to; hence, they do not actually represent linear transformations.

If one wishes to find a faithful linear representation of $CO(3, 1)$ then one must add two dimensions to $\mathbb{R}^4$ and use *hexaspherical coordinates* ([8]). A point of $\mathbb{R}^6$, when regarded this way, represents a pair of timelike hyperboloids in Minkowski space of the general form:

$$c_2(t - t_0)^2 - (x - x_0)^2 - (y - y_0)^2 - (z - z_0)^2 = a^2 , \tag{3.22}$$

in which $(ct_0, x_0, y_0, z_0)$ is the center of a translated light cone and $a \in \mathbb{R}$. So far, we have accounted for five coordinates, namely, $(t_0, x_0, y_0, z_0, a)$. Hence, we could use each point of $\mathbb{R}^5$ to represent a unique pair of timelike hyperboloids in Minkowski space. In order to justify the addition of a further coordinate, one notes that both sides of (3.22) are homogeneous of degree two, so an scalar multiple of a point that is on such a hyperboloid will also be on it. Hence, one can introduce a sixth coordinate by regarding the points of $\mathbb{R}^5$ as the inhomogeneous coordinates of $\mathbb{R}P^5$ and the points of $\mathbb{R}^6$ as homogeneous coordinates. In effect, the pairs of timelike hyperboloids in Minkowski space play the same role in conformal Lorentz geometry that the 3-spheres of radius $a$ and center $(ct_0, x_0, y_0, z_0)$ would in the conformal Euclidian geometry of $\mathbb{R}^4$.

Although we shall go into the details at the moment, the latter extension to hexaspherical coordinates allows one faithfully represent $CO(3, 1)$ as the subgroup $SO(4, 2)$ in $GL(6; \mathbb{R})$.

One also notes that if one wishes to start with $GL(4; \mathbb{R})$ and look for subgroups that pertain to conformal geometry, just as $SL(4; \mathbb{R})$ pertains to three-dimensional projective geometry, then one need only subtract two from the dimension and note that the subgroup $SO(3, 1)$ gives a faithful linear representation of a finite-dimensional subgroup of the two-dimensional Euclidian conformal group. As is widely known at present, the dimension two is distinguished in conformal Euclidian geometry as the only dimension in which the conformal group is infinite-dimensional, since one can regard compactified $\mathbb{R}^2$ as the Riemann sphere that is associated with compactifying $\mathbb{C}$, in which case, all analytic transformations of $\mathbb{C}$ become conformal transformations.

If one follows this Ansatz then the points of $\mathbb{R}^4$ are the quadrispherical coordinates of circles in $S^2$, which is regarded as compactified $\mathbb{R}^2$. Although the 2-sphere in question can be regarded abstractly, one can also associate it with the light sphere in the affine subspace $\Pi_3$. The action of $SO(3, 1)$ on this 2-sphere by conformal transformations then leads to a deeper geometric interpretation of the methods of 2-spinors, which now briefly mention.

---

[8] More generally, one simply refers to "polyspherical" coordinates.

### 3.6 Projective geometry and spinors

It is well known by both mathematicians and physicists that the connected component of the identity in the Lorentz group – viz., $SO_0(3, 1)$ – can be covered in a two-to-one way by the group $SL(2; \mathbb{C})$. Since we have just discussed the fact that the group $SL(n+1; \mathbb{R})$ can be regarded as the group of projective transformations of $\mathbb{R}P^n$, we need only point out that the basic machinery of projective geometry can be – and usually is – generalized to a less specific field of scalars than $\mathbb{R}$. In particular, the same considerations apply to the field $\mathbb{C}$.

This says that $SL(2; \mathbb{C})$ is the group of projective transformations of the projective space $\mathbb{C}P^1$, which is diffeomorphic to the real two-sphere $S^2$, which is more conveniently represented as the complex line compactified by a point at infinity into the Gaussian sphere. The association of points on the Gaussian sphere with points on the complex line is then by stereographic projection.

The complex vector space $\mathbb{C}^2 - \{0\}$ can be regarded as the space of complex homogeneous coordinates $(\xi, \zeta)$ for any point on the 2-sphere, whereas the complex number $z = \xi/\zeta$ $(\zeta \neq 0)$ can be regarded as its inhomogeneous coordinate. Hence, the action of $SL(2; \mathbb{C})$ on $\mathbb{C}^2 - \{0\}$ is simply the defining representation and the action of $\mathbb{C}^2 - \{0\}$ on $S^2$ is by *Möbius transformations* (cf., Schwerdtfegger [**16**]), which are what linear fractional transformations are referred to in the present set of circumstances. A representation of $SO_0(3, 1)$ in $SL(2; \mathbb{C})$ would similarly define an action of $SO_0(3, 1)$ on both spaces, *a fortiori*.

To get back to physics, the vectors in the space $\mathbb{C}^2$ are referred to as *spin vectors*. The extension from wave functions with values in $\mathbb{C}$ to wave functions with values in $\mathbb{C}^2$ was Pauli's way of accounting for the spin of the electron in the Schrödinger equation. Hence, in one sense wave functions with values in $\mathbb{C}^2$ – i.e., *2-spinor fields* – are still basically a non-relativistic device in physics, despite the fact that there is a natural action of $SL(2; \mathbb{C})$ on them. However, they are still useful in the electromagnetism, which we shall discuss shortly.

Now, if one identifies the two-sphere, hence, $\mathbb{C}$, with the (future) light sphere in the affine subspace $\Pi_3$, which has been given a Euclidean metric then one sees that 2-spinor wave functions effectively take their values in this light sphere and that the action of the Lorentz group on them – by means of a representation in $SL(2; \mathbb{C})$ – is also an action of the Lorentz group on this sphere.

We shall not go further in this direction at the moment, except to refer the ambitious to the literature (cf., Penrose and Rindler [**17**].).

## 4 Projective geometry and electromagnetism

### 4.1 The space of bivectors on $\mathbb{R}^4$

A bivector is, by definition, an anti-symmetric second rank contravariant tensor over a vector space; in the present case, that vector space will be $\mathbb{R}^4$. Hence, a bivector on $\mathbb{R}^4$ will be an element of the six-dimensional vector space $\Lambda_2(\mathbb{R}^4)$.

In order to define a convenient class of bases – i.e., 6-frames – for $\Lambda_2(\mathbb{R}^4)$ we first point out that when $\mathbb{R}^4$ has been given a 1+3 decomposition $[\mathbf{t}] \oplus \Pi_3$, which is associated with the pair $([\mathbf{t}], [\theta]) \in \mathbb{R}P^3 \times \mathbb{R}P^{3*}$, one can define an injection of $\Pi_3$ into $\Lambda_2(\mathbb{R}^4)$ by way of:

$$\iota_{\mathbf{t}}: \Pi_3 \to \Lambda_2(\mathbb{R}^4), \qquad \mathbf{v} \mapsto \mathbf{t} \wedge \mathbf{v} . \tag{4.1}$$

Because this map a linear injection, its image $\Lambda_{\mathrm{Re}}(\mathbb{R}^4)$ is a three-dimensional subspace of $\Lambda_2(\mathbb{R}^4)$ and this map takes a 3-frame $\mathbf{e}_i$, $i = 1, 2, 3$ in $\Pi_3$ to a 3-frame $\mathbf{E}_i \equiv \mathbf{t} \wedge \mathbf{e}_i$ in $\Lambda_2(\mathbb{R}^4)$. In order to complete this 3-frame to a 6-frame, we note the fact, which we will not prove here (cf., Delphenich [**16**]), that $\Lambda_2(\Pi_3)$ – i.e., the space of bivectors over $\Pi_3$ – is a three-dimensional subspace $\Lambda_{\mathrm{Im}}(\mathbb{R}^4)$ of $\Lambda_2(\mathbb{R}^4)$ that is complementary to the space $\Lambda_{\mathrm{Re}}(\mathbb{R}^4)$. Hence, $\Lambda_2(\mathbb{R}^4) = \Lambda_{\mathrm{Re}}(\mathbb{R}^4) \oplus \Lambda_{\mathrm{Im}}(\mathbb{R}^4)$. Since a basis for $\Lambda_2(\Pi_3)$ can be defined by the three bivectors $\mathbf{E}_{i+3} \equiv \varepsilon_{ijk}\, \mathbf{e}_j \wedge \mathbf{e}_k = \{\mathbf{e}_2 \wedge \mathbf{e}_3, \mathbf{e}_3 \wedge \mathbf{e}_1, \mathbf{e}_1 \wedge \mathbf{e}_2\}$, we have associated our 4-frame $\{\mathbf{t}, \mathbf{e}_i\}$ with a 6-frame $\mathbf{E}_I$, I = 1, …, 6. An elementary dimension count shows that not every 6-frame can be so represented, but it is intriguing that there is, nevertheless, a one-to-one correspondence between oriented time-oriented *Lorentz* 4-frames and $SO(3; \mathrm{C})$ frames in $\Lambda_2(\mathbb{R}^4)$, when it is given a complex structure and a complex orthogonal structure, which we now discuss.

In the last paragraph, we suggestively labeled our complementary 3-planes in $\Lambda_2(\mathbb{R}^4)$ as if they were, in some sense, "real" and "imaginary." We now justify this. We could simply use the 6-frame that we defined in order to define a complex structure on $\Lambda_2(\mathbb{R}^4)$, viz., a linear isomorphism $*: \Lambda_2(\mathbb{R}^4) \to \Lambda_2(\mathbb{R}^4)$ such that $*^2 = -I$:

$$*E_i \equiv E_{i+3}, \qquad *E_{i+3} \equiv -E_i . \tag{4.2}$$

and extend by linearity to the rest of $\Lambda_2(\mathbb{R}^4)$. However, in pre-metric electromagnetism [**1-5, 18-20**] the introduction of a complex structure is a consequence of having defined both a unit-volume element on $T(\mathbb{R}^4)$ and a linear electromagnetic constitutive law on $\Lambda_2(\mathbb{R}^4)$. Hence, one must understand that the complex structure is usually introduced more axiomatically.

If we use the electromagnetic notation for an arbitrary bivector $\mathfrak{F}$ then when one is given a 1+3 decomposition of $\mathbb{R}^4$ one can represent $\mathfrak{F}$ as:

$$\mathfrak{F} = \mathbf{t} \wedge \mathbf{E} - *(\mathbf{t} \wedge \mathbf{B}) = E^i\, \mathbf{E}_i - B^i\, *\mathbf{E}_i . \tag{4.3}$$

In order to exhibit an isomorphism of $\Lambda_2(\mathbb{R}^4)$ with $\mathbb{C}^3$, we first need to define the action of complex scalars on $\Lambda_2(\mathbb{R}^4)$. All one need to do is define:

$$i\mathbf{a} \equiv *\mathbf{a}, \qquad \mathbf{a} \in \Lambda_2(\mathbb{R}^4). \tag{4.4}$$

The action of a more general complex number is then defined by:

$$(\alpha + i\beta)\mathbf{a} = \alpha\mathbf{a} + \beta *\mathbf{a} . \tag{4.5}$$

As a result of this definition, and the fact that we have defined our $\mathbf{E}_{i+3} = *\mathbf{E}_i = i\mathbf{E}_i$, we see that $\{\mathbf{E}_i, i = 1, 2, 3\}$ also defines a frame for the *complex* three-dimensional vector space $\Lambda_2(\mathbb{R}^4)$, as well as a choice of isomorphism with $\mathbb{C}^3$. The isomorphism is simply the one that takes $\mathbf{a} = a^i\mathbf{E}_i$ to the complex triple $(a^1, a^2, a^3)$; i.e., the complex 3-frame $\mathbf{E}_i$ on $\Lambda_2(\mathbb{R}^4)$ goes to the canonical complex 3-frame on $\mathbb{C}^3$. Under this isomorphism, we do indeed have that $\Lambda_{\mathrm{Re}}(\mathbb{R}^4)$ goes to the real subspace of $\mathbb{C}^3$ that consists of triples of real numbers and $\Lambda_{\mathrm{Im}}(\mathbb{R}^4)$ goes to the subspace of imaginary triples.

With the aforementioned electromagnetic notation, we see that the complex components of an arbitrary $\mathfrak{F}$, $*\mathfrak{F} \in \Lambda_2(\mathbb{R}^4)$ become:

$$\mathfrak{F}^i = E^i - iB^i, \quad *\mathfrak{F}^i = B^i + iE^i = i\mathfrak{F}^i. \tag{4.6}$$

When one introduces a unit-volume element $\varepsilon \in \Lambda^4(\mathbb{R}^4)$ on $\mathbb{R}^4$ (really, on its tangent bundle), one can define a real scalar product of signature type (3, 3) on $\Lambda_2(\mathbb{R}^4)$ by way of:

$$<\mathbf{a}, \mathbf{b}> = \varepsilon(\mathbf{a} \wedge \mathbf{b}) = <\mathbf{a}_{\mathrm{R}}, \mathbf{b}_{\mathrm{I}}> + <\mathbf{a}_{\mathrm{I}}, \mathbf{b}_{\mathrm{R}}>. \tag{4.7}$$

In this expression, we have used the subscripts R and I to indicate the real and imaginary components of the bivectors relative to the chosen decomposition of $\Lambda_2(\mathbb{R}^4)$.

This scalar product defines a quadratic form by way of:

$$Q(\mathbf{a}) = <\mathbf{a}, \mathbf{a}> = \varepsilon(\mathbf{a} \wedge \mathbf{a}) = 2 <\mathbf{a}_{\mathrm{R}}, \mathbf{a}_{\mathrm{I}}>. \tag{4.8}$$

When one now introduces a complex structure * on $\Lambda_2(\mathbb{R}^4)$, as well, one can not only introduce a second real scalar product on $\Lambda_2(\mathbb{R}^4)$:

$$(\mathbf{a}, \mathbf{b}) = \varepsilon(\mathbf{a} \wedge *\mathbf{b}) = (\mathbf{a}_{\mathrm{R}}, \mathbf{b}_{\mathrm{R}}) - (\mathbf{a}_{\mathrm{I}}, \mathbf{b}_{\mathrm{I}}) , \tag{4.9}$$

which is also of signature type (3, 3) and has the quadratic form:

$$Q^*(\mathbf{a}) = (\mathbf{a}, \mathbf{a}) = \varepsilon(\mathbf{a} \wedge *\mathbf{a}) = (\mathbf{a}_{\mathrm{R}}, \mathbf{a}_{\mathrm{R}}) - (\mathbf{a}_{\mathrm{I}}, \mathbf{a}_{\mathrm{I}}), \tag{4.10}$$

but also a complex scalar product:

$$<\mathbf{a}, \mathbf{b}>_{\mathbb{C}} = <\mathbf{a}, \mathbf{b}> + i(\mathbf{a}, \mathbf{b}) . \tag{4.11}$$

A complex 3-frame on $\Lambda_2(\mathbb{R}^4)$, such as $\mathbf{E}_i$, is called *complex orthogonal* iff $<\mathbf{E}_i, \mathbf{E}_j>_{\mathbb{C}} = \delta_{ij}$ . The association of $\mathbf{E}_i$ with the canonical 3-frame on $\mathbb{C}^3$ defines not only a $\mathbb{C}$-linear isomorphism of the two vector spaces, but also an isometry of the two complex orthogonal spaces when $\mathbb{C}^3$ is given the complex Euclidian structure, whose components relative to that canonical 3-frame are $\delta_{ij}$.

### 4.2 The physical nature of real and imaginary 3-planes in $\Lambda_2(\mathbb{R}^4)$

Although the geometry of $\Lambda_2(\mathbb{R}^4)$ seems somewhat esoteric by physical standards to the uninitiated, nevertheless, it is possible to regard it as a natural setting for the description of relative motions in space, and even casts a more subtle light on the conceptual nature of the introduction of a fourth dimension in the description of physical motions.

The essential ideas center around the injection of 3-planes in Minkowski space into $\Lambda_2(\mathbb{R}^4)$ by way of lines not contained in those 3-planes. As we have emphasized, the decomposition of $\Lambda_2(\mathbb{R}^4)$ into $\Lambda_{Re}(\mathbb{R}^4) \oplus \Lambda_{Im}(\mathbb{R}^4)$ is therefore closely related to the decomposition of $\mathbb{R}^4$ into $[\mathbf{t}] \oplus \Pi_3$. In the context of electromagnetism the decomposition of $\Lambda_2(\mathbb{R}^4)$ is a decomposition into electric and magnetic subspaces, which is also equivalent to a choice of rest frame in Minkowski space.

Hence, one can just as well regard the decomposition of $\Lambda_2(\mathbb{R}^4)$ as being a decomposition into a "position space" and a "boost space." This is entirely consistent with the aforementioned ambiguity associated with trying to extend the injection of $\Pi_3$ by $[\mathbf{t}]$ to an injection of $\mathbb{R}^4$, namely, where one is to put $[\mathbf{t}]$ in the imaginary subspace $*([\mathbf{t}] \wedge \Pi_3)$. Since this ambiguity is simply the ambiguity of where to put a boost direction in $\Pi_3$, this is consistent with our decision to regard $\Lambda_{Im}(\mathbb{R}^4) = *([\mathbf{t}] \wedge \Pi_3) = \Lambda_2(\Pi_3)$ as a sort of boost space that is dual to the position space $\Pi_3$.

Of course, this decomposition of the six-dimensional vector space $\Lambda_2(\mathbb{R}^4)$ is also closely analogous to the decomposition of the six-dimensional vector space that underlies $\mathfrak{so}(3, 1)$ into the direct sum of $\mathfrak{so}(3)$ and the space of infinitesimal boosts, which is also isomorphic to the decomposition of $\mathfrak{so}(3; \mathbb{C})$ into real and imaginary infinitesimal rotations. It is this latter decomposition that is most closely analogous to a decomposition of $\Lambda_2(\mathbb{R}^4)$ into real and imaginary subspaces.

It is important to note, as we have, that such a choice of decomposition of $\Lambda_2(\mathbb{R}^4)$ is not unique, and since we related such decompositions to the corresponding decomposition of $\mathbb{R}^4$ into time and space, one sees that a decomposition of $\Lambda_2(\mathbb{R}^4)$ implies that one respect the same sort of equivalence of all such choices. We shall see that the group of linear transformations of $\Lambda_2(\mathbb{R}^4)$ that takes oriented decompositions to other oriented decompositions is isomorphic to $SL(3; \mathbb{C})$, which is also the group of projective transformations of $\mathbb{R}P^2$, and includes the subgroup $SL(2; \mathbb{C})$, which is fundamental to the spinor representations of the proper orthochronous Lorentz group.

What we shall try to establish in the next section is that the physically significant point this that cannot be over-emphasized is that the origin of the Lorentzian structure on $\mathbb{R}^4$, which is traditionally associated with the presence of gravitation in spacetime, seems to be more fundamentally related to the presence of electromagnetism, or rather, the electromagnetic constitutive properties of the vacuum state that facilitate electromagnetic wave motion. In effect, in the simplest case of linear isotropic homogeneous constitutive laws one needs to first regard $c_0^2$ as being more essentially described by $1/\varepsilon_0\mu_0$ and then replace the Lorentzian metric of the form $c_0^2 dt^2 - dx^2 - dy^2 - dz^2$ with its "electromagnetic" form $\mu_0^{-1} dt^2 - \varepsilon_0(dx^2 - dy^2 - dz^2)$. One must note that although the constancy of $c_0$ under Lorentz transformations is generally axiomatic, the constancy of $\varepsilon_0$ and $\mu_0$ is not.

One can then pursue the consequences of weakening the assumptions concerning the electromagnetic constitutive properties of the vacuum state. In particular, a particularly compelling direction seems to be that of introducing vacuum polarization at high field strengths, which would introduce nonlinearity into the electromagnetic constitutive properties.

### 4.3 Decomposable 2-forms as elementary electromagnetic fields

For the sake of physical familiarity, we now switch from discussing bivectors to discussing 2-forms; of course, the concepts that we will be discussing are equally applicable to either space.

So far, we have been discussing 2-forms of the general form $F = \theta \wedge E - *(\theta \wedge B)$. A non-zero 2-form $F \in \Lambda^2(\mathbb{R}^4)$ is called *decomposable* if it can be expressed in the form:

$$F = \alpha \wedge \beta . \tag{4.12}$$

This representation is not unique, since if one subjects the linearly independent 1-forms $\alpha$ and $\beta$ to a general linear transformation in the plane that they span then the 2-form $F$ is multiplied by the determinant of that transformation. Hence, if that determinant is one then $F$ remains unchanged. Furthermore, $F$ can also be represented as a sum of two terms of the form (4.12) by substituting $\alpha = \gamma + \delta$, or even more, if one also substitutes similarly for $\beta$.

A basic property of decomposable 2-forms is that they all satisfy $F \wedge F = 0$; i.e., the annihilating 2-plane that $F$ defines obviously has a non-trivial intersection with itself. Therefore, one can also say that:

$$Q(F) = \langle F, F\rangle = 0 . \tag{4.13}$$

Hence, the set of all decomposable $F$ is a quadric surface in $\Lambda^2(\mathbb{R}^4)$ that one calls the *Klein quadric*. Since equation (4.13) is also homogeneous of degree two, one also defines a quadric in the projective space that is associated with the vector space $\Lambda^2(\mathbb{R}^4)$, which is non-canonically diffeomorphic to $\mathbb{R}P^5$.

Decomposable 2-forms on $\mathbb{R}^4$ play a special role both in geometry and in physics. In projective geometry, they represent 2-planes in $\mathbb{R}^4$, or its dual by way of the *Plücker embedding*, either by associating the line through $F$ with the 2-plane in $\mathbb{R}^{4*}$ that $\alpha$ and $\beta$ span directly, or the 2-plane in $\mathbb{R}^4$ that the line through $F$ annihilates.

In physics, one notes that in the electromagnetic representation of a general 2-form, there are two obvious cases in which $F$ is decomposable:

*i*) The *purely electric* case (relative to the choice of $(\mathbf{t}, \theta)$):

$$F = \theta \wedge E. \tag{4.14}$$

*ii*) The *purely magnetic* case:

$$F = -*(\theta \wedge B) . \tag{4.15}$$

Although this does not appear to have the required form, it helps to known that * preserves the scalar product <.,.>.

There is also a third case that is less immediately apparent:

*iii*) The *isotropic* case:

$$F = l \wedge E = (\theta + n) \wedge E = \theta \wedge E + n \wedge E, \tag{4.16}$$

in which the 1-form $l$ is lightlike and $n$ is spacelike.

If we use the complex scalar product $<.,.>_{\mathbb{C}}$ on $\Lambda^2(\mathbb{R}^4)$ that is defined by * and $\varepsilon$ then since decomposable 2-forms will be purely imaginary and $<F, F>_{\mathbb{C}} = i(E^2 - B^2)$ we see that the three cases can be classified by the sign of the imaginary part of $<F, F>_{\mathbb{C}}$: $F$ is purely electric when it is positive, isotropic when it is zero, and purely magnetic when it is negative.

Since time-varying $E$ or $B$ fields will induce time-varying $B$ or $E$ fields, respectively, the purely electric or magnetic 2-forms must be static fields, at least in the rest space defined by θ. Although electromagnetic waves are represented by isotropic 2-forms, the reason that we hesitated to use the term "lightlike" for isotropic 2-forms is that one can also configure superpositions of static $E$ and $B$ fields that define isotropic 2-forms; all one needs to do is make them perpendicular with their magnitudes equal. However, since physical fields in reality always must have sources, one speculates that the essential difference between decomposable 2-forms and the more general ones is that the general fields cannot have a single source. Hence, we regard the decomposable 2-forms as essentially the elementary building blocks of more complicated electromagnetic field configurations.

It is worth noting that since the set of decomposable 2-forms is a closed five-dimensional subset of a six-dimensional vector space, decomposable 2-forms are not the generic case. Generically, "most" 2-forms have rank four, and therefore represent superpositions of electromagnetic fields with differing sources.

### 4.4 The action of $GL(3; \mathbb{C})$ on $\Lambda_2(\mathbb{R}^4)$

Now that we have a complex structure on $\Lambda_2(\mathbb{R}^4)$ and an action of the complex scalars that makes it isomorphic to $\mathbb{C}^3$, it makes sense to first investigate the subgroup of $GL(\Lambda_2(\mathbb{R}^4))$, which is non-canonically isomorphic to $GL(6; \mathbb{R})$, that preserves the complex structure.

The condition for this is that if $C \in GL(6; \mathbb{R})$ preserves * then $C$ must satisfy:

$$C*C^{-1} = *, \tag{4.17a}$$

or, equivalently:

$$C^* = {}^*C\,. \tag{4.17b}$$

Hence, $C$ must commute with *.

A straightforward, but tedious, computation using a general matrix $C$ partitioned into block form gives that $C$ must be of the form:

$$C = \left[\begin{array}{c|c} A & -B \\ \hline B & A \end{array}\right] = \left[\begin{array}{c|c} A & 0 \\ \hline 0 & A \end{array}\right] + {}^*\left[\begin{array}{c|c} B & 0 \\ \hline 0 & B \end{array}\right], \tag{4.18}$$

in which $A, B \in GL(3;\mathbb{R})$. We can associate any such $C$ with a unique element of $GL(3;\mathbb{C})$ by way of the assignment:

$$GL(3;\mathbb{C}) \to GL(6;\mathbb{R}), \qquad A + iB \mapsto A + {}^*B, \tag{4.19}$$

which is also a group isomorphism onto. Hence, the subgroup of $GL(6;\mathbb{R})$ that preserves the complex structure defined by * is isomorphic to $GL(3;\mathbb{R})$.

We now look at the projective spaces that are associated with $\Lambda_2(\mathbb{R}^4)$. Actually, there are two projectivizations of $\Lambda_2(\mathbb{R}^4)$ that we can discuss: the real projectivization and the complex one.

In the latter case, we are speaking of the space $P\Lambda_2(\mathbb{R}^4)$ of all real lines through the origin of $\Lambda_2(\mathbb{R}^4)$, a five-dimensional real manifold that is diffeomorphic to $\mathbb{R}P^5$; as we saw, it is the proper setting for the Klein quadric.

However, since it is becoming apparent that it is the complex structure that plays the fundamental role in the geometry of $\Lambda_2(\mathbb{R}^4)$, we should really focus on the space $\mathbb{C}P(\Lambda_2(\mathbb{R}^4))$ of all *complex* lines through the origin of $\Lambda_2(\mathbb{R}^4)$, a 2-dimensional complex manifold that is biholomorphic to $\mathbb{C}P^2$, which may also be regarded as a four-dimensional real manifold.

In order to get more intuition for the notion of a complex line through the origin in $\Lambda_2(\mathbb{R}^4)$, let us first consider a complex line $[z^i]$ through the origin of $\mathbb{C}^3$ and see what it maps to under the isomorphism that was defined previously. If $z^i = x^i + iy^i$ then $[z]$ is the equivalence class of all complex numbers of the form: $\lambda z^i = (\alpha + i\beta)\, z^i = \alpha z^i + i\beta z^i$.

Hence, under the isomorphism above, if $z^i$ corresponds to the bivector $\boldsymbol{\alpha}$ then the complex line through the origin of $\Lambda_2(\mathbb{R}^4)$ that it generates, which we denote by $[\boldsymbol{\alpha}]$, consists of the two-dimensional subspace of all bivectors of the form: $\alpha\boldsymbol{\alpha} + \beta\, {}^*\boldsymbol{\alpha}$. Since this is the 2-plane spanned by $\boldsymbol{\alpha}$ and ${}^*\boldsymbol{\alpha}$, we see that the fundamental objects of complex projective geometry in $\Lambda_2(\mathbb{R}^4)$ are the *duality planes* that are associated with each bivector.

The two complex dimensions that parameterize a chart of $\mathbb{C}P(\Lambda_2(\mathbb{R}^4))$ can be described by the complex inhomogeneous coordinates that one obtains by projecting from $(z^1, z^2, z^3)$ to $(Z^1, Z^2)$ when $z^1 \neq 0$, where $Z^1 = z^2/z^1$, $Z^2 = z^3/z^1$. These coordinates describe a complex affine 2-plane in $\mathbb{C}P(\Lambda_2(\mathbb{R}^4))$ – hence, a real affine 4-plane – that each complex line – i.e., duality plane – intersects exactly once. One can characterize it as the complex affine 2-plane in $\mathbb{C}^3$ that is defined by $z^1 = 1$, which corresponds to the real affine 4-plane in $\Lambda_2(\mathbb{R}^4)$ that consists of all bivectors of the form:

$$\mathbf{F} = \mathbf{E}_1 + E^i\,\mathbf{E}_i + B^i\,{*}\mathbf{E}_i\,, \quad i = 2, 3. \tag{4.20}$$

Of course, there is nothing special about the choice of $z^1$, so there are corresponding affine 4-planes associated with choosing $z^2$ or $z^3$, for instance.

If we regard the space $\Lambda_2(\mathbb{R}^4)$ – or rather its isomorphic copy $\mathbb{C}^3$ – as something like the complex homogeneous coordinates of the elements of $\mathbb{C}P\Lambda_2(\mathbb{R}^4)$ then if we can introduce a choice of unit-volume element on $\Lambda_2(\mathbb{R}^4)$, we can reduce the action of $GL(3; \mathbb{C})$ on $\Lambda_2(\mathbb{R}^4)$ to an action of $SL(3; \mathbb{C})$, which then induces homogeneous and inhomogeneous projective transformations of the complex lines in $\Lambda_2(\mathbb{R}^4)$. Such a unit-volume element is easy to find, since $\Lambda_2(\mathbb{R}^4)$, like any vector space, is orientable. It would represent a non-vanishing 6-vector $\mathcal{V}_\Lambda$ defined in the exterior algebra of the six-dimensional vector space $\Lambda_2(\mathbb{R}^4)$, which is to be distinguished from the exterior algebra of the four-dimensional vector space $\mathbb{R}^4$, in which 6-vectors would necessarily vanish.

A convenient choice of unit-volume element is given by:

$$\mathcal{V}_\Lambda = E^1 \perp E^2 \perp E^3 = \frac{1}{3!}\,\varepsilon_{ijk}\, E^i \perp E^j \perp E^k, \tag{4.21}$$

which can be also be regarded as a complex 3-form over $\mathbb{C}^{3*}$. Here, we are using the symbol $\perp$ to distinguish the exterior product over the vector space $\Lambda_2(\mathbb{R}^4)$ from the exterior product over the vector space $\mathbb{R}^4$.

We can then reduce the action of $GL(3; \mathbb{C})$ on $\Lambda_2(\mathbb{R}^4)$ to an action of $SL(3; \mathbb{C})$ by restricting ourselves to the elements of $GL(3; \mathbb{C})$ that preserve $\mathcal{V}_\Lambda$. The real 6×6 matrix form of such transformations for a given choice of complex 6-frame on $\Lambda_2(\mathbb{R}^4)$ is then still of the form (4.18), except that now its determinant must be one, which implies that:

$$\det(A)^2 + \det(B)^2 = 1. \tag{4.22}$$

The elements of $SL(3; \mathbb{C})$ can then be described as before in terms of their action on the complex affine 2-plane $z^0 = 1$. The complex and real forms of the matrices are:

*Homotheties:*

$$\left[\begin{array}{c|cc} C_0^0 & 0 & 0 \\ \hline 0 & 1 & 0 \\ 0 & 0 & 1 \end{array}\right] = \left[\begin{array}{ccc|ccc} A_0^0 & 0 & 0 & B_0^0 & 0 & 0 \\ 0 & 1 & 0 & 0 & 1 & 0 \\ 0 & 0 & 1 & 0 & 0 & 1 \\ \hline -B_0^0 & 0 & 0 & A_0^0 & 0 & 0 \\ 0 & -1 & 0 & 0 & 1 & 0 \\ 0 & 0 & -1 & 0 & 0 & 1 \end{array}\right] \tag{4.23a}$$

$$\textit{Translations:}\qquad \left[\begin{array}{c|cc} 1 & 0 & 0 \\ \hline C_0^1 & 1 & 0 \\ C_0^2 & 0 & 1 \end{array}\right] = \left[\begin{array}{ccc|ccc} 1 & 0 & 0 & 1 & 0 & 0 \\ A_0^1 & 1 & 0 & B_0^1 & 1 & 0 \\ A_0^2 & 0 & 1 & B_0^2 & 0 & 1 \\ \hline -1 & 0 & 0 & 1 & 0 & 0 \\ -B_0^1 & -1 & 0 & A_0^1 & 1 & 0 \\ -B_0^2 & 0 & -1 & A_0^2 & 0 & 1 \end{array}\right] \tag{4.23b}$$

$$\textit{Inversions:}\qquad \left[\begin{array}{c|cc} 1 & C_1^0 & C_2^0 \\ \hline 0 & 1 & 0 \\ 0 & 0 & 1 \end{array}\right] = \left[\begin{array}{ccc|ccc} 1 & A_1^0 & A_2^0 & 1 & B_1^0 & B_2^0 \\ 0 & 1 & 0 & 0 & 1 & 0 \\ 0 & 0 & 1 & 0 & 0 & 0 \\ \hline -1 & -B_1^0 & -B_2^0 & 1 & A_1^0 & A_2^1 \\ 0 & -1 & 0 & 0 & 1 & 0 \\ 0 & 0 & -1 & 0 & 0 & 1 \end{array}\right] \tag{4.23c}$$

$$\textit{Linear transformations:}\qquad \left[\begin{array}{c|cc} 1 & 0 & 0 \\ \hline 0 & C_1^1 & C_2^1 \\ 0 & C_1^2 & C_2^2 \end{array}\right] = \left[\begin{array}{ccc|ccc} 1 & 0 & 0 & 1 & 0 & 0 \\ 0 & A_1^1 & A_2^1 & 0 & B_1^1 & B_2^1 \\ 0 & A_1^2 & A_2^2 & 0 & B_2^1 & B_2^2 \\ \hline -1 & 0 & 0 & 1 & 0 & 0 \\ 0 & -B_1^1 & -B_2^1 & 0 & A_1^1 & A_2^1 \\ 0 & -B_1^2 & -B_2^2 & 0 & A_1^2 & A_2^2 \end{array}\right], \tag{4.23d}$$

in which we have set $C_j^i = A_j^i + iB_j^i$.

The action of $SL(3; \mathbb{C})$ on the inhomogeneous coordinates of points in the complex affine 2-plane is then:

$$C_j^i\,[Z^1, Z^2]^{\mathrm{T}} = \left[\frac{C_0^1 + C_i^1 Z^i}{C_0^0 + C_i^0 Z^i},\quad \frac{C_0^2 + C_i^2 Z^i}{C_0^0 + C_i^0 Z^i}\right]^{\mathrm{T}}. \tag{4.24}$$

As we pointed out above, the complex vector space $\Lambda_2(\mathbb{R}^4)$ can be given a complex orthogonal structure, as we defined in (4.11). Hence, one can further reduce the group $SL(3; \mathbb{C})$ to the subgroup of invertible complex transformations that not only preserve the unit-volume element, but also the complex orthogonal structure, which gives a subgroup of $GL(6; \mathbb{R})$ that is isomorphic to not only $SO(3; \mathbb{C})$, but also $SO_0(3, 1)$.

It is illuminating to observe that when Lorentz transformations are represented by elements of $SO(3; \mathbb{C})$, the difference between Euclidian rotations and Lorentz boosts – i.e., hyperbolic rotations – is only a matter of a multiplication by $i$. One simply expresses the complex 3×3 orthogonal matrix as a sum of a real 3×3 orthogonal matrix plus $i$ times another such matrix. This also shows why the product of two elementary boosts is a rotation: because $i^2$ is a real number. Of course, one can also regard hyperbolic rotations as rotations through $i$ times a real angle.

If one regards an element of $SO(3; \mathbb{C})$ as a 6×6 real matrix of the form (4.18), and the space of bivectors as a direct sum of an electric and a magnetic subspace then one sees that the effect of the Euclidian rotations is to take electric bivectors to electric bivectors and magnetic ones to magnetic ones. Since a hyperbolic rotation is the product of a Euclidian rotation with *, one need only note that * then takes an electric bivector to a magnetic one and a magnetic bivector to minus an electric one.

To summarize the relationships between $\mathbb{R}^4$, the space of bivectors over $\mathbb{R}^4$, and $\mathbb{C}^3$, we state the following "pseudo-theorem":

TFARE ([9]):

*i*. Time + space decompositions of $\mathbb{R}^4$.
*ii*. Electric + magnetic decompositions of $\Lambda_2(\mathbb{R}^4)$.
*iii*. Real + imaginary decompositions of $\mathbb{C}^3$.

More rigorous details along these lines are found in Delphenich [**18**].

There is a second way of reducing from $SL(3; \mathbb{C})$ to a subgroup that is (two-to-one) isomorphic to $SO_0(3, 1)$ – namely, $SL(2; \mathbb{C})$ – that does not directly involve imposing a complex Euclidian structure on $\Lambda_2(\mathbb{R}^4)$. Indeed, one sees that the group $SL(2; \mathbb{C})$ is more intrinsically defined as the group of projective transformations of $\mathbb{C}P^1$, which act on the homogeneous coordinates in $\mathbb{C}^2 - \{0\}$ by matrix multiplication and on the inhomogeneous coordinates of $\mathbb{C}P^1$ by linear fractional transformations. As mentioned above, since $\mathbb{C}P^1$ is really just $S^2$, which one envisions to be the one-point compactification of the complex line (regarded as a real 2-plane), in this particular case the linear fractional transformations are simply the Möbius transformations.

In order to specify the reduction of $SL(3; \mathbb{C})$ to $SL(2; \mathbb{C})$, note that since the determinant of any element $A \in SL(3; \mathbb{C})$ is one, if $\lambda_i$, $i = 1, 2, 3$ are the eigenvalues of $A$ then one must have $\lambda_1 \lambda_2 \lambda_3 = 1$. Similarly, if $\mu_1\ \mu_2$ are eigenvalues of an element $B \in SL(2; \mathbb{C})$ then one must have $\mu_1 \mu_2 = 1$. Hence, if $B$ is represented by an element of $SL(3; \mathbb{C})$ such that $\mu_1 = \lambda_1$, $\mu_2 = \lambda_2$, one must have that $\lambda_3 = 1$. Hence, any element of an $SL(2; \mathbb{C})$ subgroup of $SL(3; \mathbb{C})$ must fix some complex line through the origin in $\mathbb{C}^3$; i.e., some element $[\mathbf{z}] \in \mathbb{C}P^2$. (Recall that the image of a complex line through origin in $\mathbb{C}^3$ is represented by a duality plane in $\Lambda_2(\mathbb{R}^4)$).

However, since the difference between the complex dimensions of $SL(3; \mathbb{C})$ and $SL(2; \mathbb{C})$ is $8 - 3 = 5$, and the complex dimension of $\mathbb{C}P^2$ is two, one also sees that merely specifying the complex line that is fixed by the elements of an $SL(2; \mathbb{C})$ subgroup is not sufficient in order to specify the subgroup geometrically, since there are three more dimensions to account for. For one thing, one must specify a complementary complex 2-plane to the fixed complex line. This will account for two more dimensions, since the space of complex 2-planes in $\mathbb{C}^3$ is also the projectivization of the space $\mathbb{C}^{3*}$ of $\mathbb{C}$-linear functionals on $\mathbb{C}^3$, which is also diffeomorphic to $\mathbb{C}P^2$. Hence, we are specifying a 1+2 decomposition of $\mathbb{C}^3$ by way of a pair of elements $([\mathbf{z}], [\psi]) \in \mathbb{C}P^2 \times \mathbb{C}P^{2*}$ such that $[\mathbf{z}]$ is not incident on $[\psi]$ ; i.e., $[\psi][\mathbf{z}] \neq 0$. The remaining dimension we need to account for is a choice of scale factor, since the orbits of the action of $SL(2; \mathbb{C})$ on all linear frames that are adapted to a given choice of ([**z**], [$\psi$]) are parameterized by their complex volumes.

[9] The Following Are Roughly Equivalent!

Now, to relate all of this back to the space of bivectors on $\mathbb{R}^4$, we note that a 1+2 decomposition of $\mathbb{C}^3$ corresponds to a 2+4 decomposition of $\Lambda_2(\mathbb{R}^4)$, when regarded as a real vector space. Hence, the element $[\mathbf{z}] \in \mathbb{C}P^2$ corresponds to a choice of duality plane and the element $[\psi] \in \mathbb{C}P^{2*}$ corresponds to a choice of complementary 4-plane. Furthermore, one notes that this complementary 4-plane must be decomposable into a complementary pair of duality planes. Finally, the choice of complex scale factor amounts a choice of not only a real scale factor, but also a choice of imaginary phase angle within the duality planes.

If we assume that we already have a 3+3 decomposition $\Pi_{\text{Re}} \oplus *\Pi_{\text{Re}}$ of $\Lambda_2(\mathbb{R}^4)$ that is consistent with * then a 2+4 decomposition can be defined by specifying a real 2+1 decomposition of either direct summand and summing the line L so defined with the other direct summand: e.g., $\Pi_{\text{Re}} \oplus \Pi_{\text{Im}} = \Pi_2 \oplus (L \oplus *\Pi_{\text{Re}}) = \Pi_2 \oplus (L \oplus *L) \oplus *\Pi_2$ . Note that the central summand $L \oplus *L$ defines a duality plane. Hence, it corresponds to a complex line through the origin in $\mathbb{C}^3$, and the remaining two summands $\Pi_2 \oplus *\Pi_2$ define a complex two-dimensional complement.

One can further refine the reduction to an $SL(2; \mathbb{C})$ subgroup by specifying the character of the line L in $\Lambda_2(\mathbb{R}^4)$: rank two or rank four, and electric, magnetic, or isotropic in the case of rank two.

One also notes that if $\Pi_{\text{Re}} = [\mathbf{t}] \wedge \Pi_3$, where $\Pi_3$ is a 3-plane in $\mathbb{R}^4$, then once one has defined a 1+2 decomposition of $\Pi_{\text{Re}}$ – hence, $\Pi_3$ – one also has the means to extend the representation of $\Pi_3$ as a 3-plane in $\Lambda_2(\mathbb{R}^4)$ to a representation of all of $\mathbb{R}^4$ as a 4-plane in $\Lambda_2(\mathbb{R}^4)$, namely, $\Pi_{\text{Re}} \oplus *L$.

To put this last statement into the electromagnetic context, we note that if the linear electromagnetic constitutive law $\chi$ on $\Lambda^2(\mathbb{R}^4)$ takes the usual homogeneous isotropic vacuum form:

$$F = E_i\,(\theta^0 \wedge \theta^i) - B_i *(\theta^0 \wedge \theta^i) \tag{4.25a}$$

$$\mathfrak{h} = \chi(F) = \varepsilon_0\, E^i\,(\mathbf{t} \wedge \mathbf{E}_i) - \frac{1}{\mu_0} B^i *(\mathbf{t} \wedge \mathbf{E}_i) \tag{4.25b}$$

then suppose we choose $\mathbf{E}_1$ to generate our line L. If we define a 4-covector by way of:

$$v = v_i\,(\theta^0 \wedge \theta^i) - v_0 *(\theta^0 \wedge \theta^1), \tag{4.26}$$

then we see that it has the property that:

$$\chi(v, v) = \chi(v)(v) = \varepsilon_0\, v^i v_i - \frac{1}{\mu_0} v^0 v_0 = \varepsilon_0\,(v^i v_i - c_0^2 v^0 v_0). \tag{4.27}$$

Hence, one sees that in this case the scalar product on $\mathbb{R}^4$ that one gets from injecting $\mathbb{R}^4$ into $\Lambda^2(\mathbb{R}^4)$ is conformal to the usual Minkowski scalar product. The essential subtlety to comprehend here is that the light cone structure of Minkowski space is still traceable to a more physically fundamental structure in the form of the linear electromagnetic constitutive law that dictates the way that electromagnetic waves will propagate. In particular, note that it is only when space is magnetically isotropic that one can choose the line L arbitrarily, since in the magnetically anisotropic case $c_0$ would generally depend upon the direction that was chosen.

### 4.5 Bivectors and bispinors

No modern discussion of electromagnetism would be complete without some mention of the role of spinors in the description of the fundamental fields of electrons/positrons and photons. Indeed, a discussion of that topic would itself be incomplete without some mention of the role that projective geometry plays.

If we introduce the Minkowski scalar product $\eta$ on $\mathbb{R}^4$ then we can define the Clifford algebra $C(\mathbb{R}^4, \eta)$ over Minkowski space most concisely by choosing an orthonormal basis $\mathbf{e}_\mu$, $\mu = 0, \ldots, 3$ , and specifying that the elements of $C(\mathbb{R}^4, \eta)$ are generated by all tensor products of the vectors $\mathbf{e}_\mu$, modulo the replacement of any expressions of the form $\mathbf{e}_\mu \otimes \mathbf{e}_\nu + \mathbf{e}_\nu \otimes \mathbf{e}_\mu$ with the scalar $2\eta_{\mu\nu}$. This defines a bilinear product on the tensor algebra over $\mathbb{R}^4$ that actually has no elements of rank higher than four, due to the aforementioned replacement. The Clifford products of the basis vectors then satisfy:

$$\mathbf{e}_\mu \mathbf{e}_\nu + \mathbf{e}_\nu \mathbf{e}_\mu = 2\eta_{\mu\nu}, \tag{4.28}$$

which is equivalent to:

$$\mathbf{e}_\mu \mathbf{e}_\nu = -\mathbf{e}_\nu \mathbf{e}_\mu, \quad \mu \neq \nu, \qquad\qquad \mathbf{e}_\mu \mathbf{e}_\mu = \eta_{\mu\mu} = \pm 1 . \tag{4.29}$$

These relations show that one can concisely express the Clifford product of two vectors in Minkowski space as:

$$\mathbf{ab} = \mathbf{a} \wedge \mathbf{b} + \eta(\mathbf{a}, \mathbf{b}). \tag{4.30}$$

Hence, **ab** agrees with $\mathbf{a} \wedge \mathbf{b}$ iff **a** is orthogonal to **b** and $\mathbf{aa} = a^2$.

Since (4.29) shows that unequal basis elements anti-commute, one sees that the Clifford algebra over Minkowski space has a lot in common with the exterior algebra over it. Indeed, if one forms all Clifford products of the orthogonal basis elements, modulo relations (4.29), one finds that there are only sixteen independent products: 1, $\mathbf{e}_\mu$, $\mathbf{e}_\mu \mathbf{e}_\nu$ $(\mu < \nu)$, $\mathbf{e}_\lambda \mathbf{e}_\mu \mathbf{e}_\nu$ $(\lambda < \mu < \nu)$, $\mathbf{e}_0 \mathbf{e}_1 \mathbf{e}_2 \mathbf{e}_3$. Hence, by associating each of these basis elements for $C(\mathbb{R}^4, \eta)$ with the corresponding exterior product, we see that there is a linear isomorphism – but not an algebra isomorphism – between $C(\mathbb{R}^4, \eta)$ and $\Lambda_*(\mathbb{R}^4)$. Indeed, one can regard $\Lambda_*(\mathbb{R}^4)$ as the Clifford algebra of $\mathbb{R}^4$ when it is given the completely degenerate scalar product $\eta = 0$.

The sixteen-dimensional Clifford algebra $C(\mathbb{R}^4, \eta)$ contains an eight-dimensional subalgebra that is spanned by just the elements 1, $\mathbf{e}_\mu \mathbf{e}_\nu$ $(\mu < \nu)$, $\mathbf{e}_0 \mathbf{e}_1 \mathbf{e}_2 \mathbf{e}_3$. Since these basis elements are all products of an even number of basis vectors, one refers to this subalgebra as the *even subalgebra* of $C(\mathbb{R}^4, \eta)$. It is, in turn, isomorphic to the Clifford algebra $C(\mathbb{R}^3, \delta)$ of $\mathbb{R}^3$ when given the Euclidian scalar product $\delta$. In fact, the isomorphism can be given by choosing the timelike basis vector $\mathbf{e}_0$ and mapping the basis vectors 1, $\mathbf{e}_i$, $i$ = 1, 2, 3, $\mathbf{e}_i\,\mathbf{e}_j$ $(i < j)$, $\mathbf{e}_1\,\mathbf{e}_2\,\mathbf{e}_3$, for $C(\mathbb{R}^3, \delta)$ to the Clifford products 1, $\mathbf{e}_0\,\mathbf{e}_i$, $\mathbf{e}_i\,\mathbf{e}_j$ $(i < j)$, $\mathbf{e}_0\,\mathbf{e}_1\,\mathbf{e}_2\,\mathbf{e}_3$ in $C(\mathbb{R}^4, \eta)$.

Note that the corresponding association of the vectors $\mathbf{e}_i$ with the bivectors $\mathbf{e}_0 \wedge \mathbf{e}_i$ was how we represented our rest space $\Pi_3$ as a subspace of $\Lambda_2(\mathbb{R}^4)$.

Under the linear isomorphism of $C(\mathbb{R}^4, \eta)$ with $\Lambda_*(\mathbb{R}^4)$, when the latter space is also given the Hodge dual operator *, one sees that these elements are represented by 1, $\mathbf{e}_0 \wedge \mathbf{e}_i$, $*(\mathbf{e}_0 \wedge \mathbf{e}_i)(i < j)$, $*1$. Since the second-degree products span the real and imaginary subspaces of $\Lambda_2(\mathbb{R}^4)$, we see that we can continue our linear isomorphism of algebras to a $\mathbb{R}$-linear isomorphism of the even subalgebra of $C(\mathbb{R}^4, \eta)$ with the complex vector space $\mathbb{C} \oplus \mathbb{C}^3 = \mathbb{C}^4$, which we regard as a real vector space. In fact, $C(\mathbb{R}^4, \eta)$ can be regarded as the complexification of $C(\mathbb{R}^3, \delta)$ under the assignment: $i(1) = \mathbf{e}_0\,\mathbf{e}_1\,\mathbf{e}_2\,\mathbf{e}_3$, $i\,\mathbf{e}_0\,\mathbf{e}_k = \varepsilon_{ijk}\,\mathbf{e}_i\,\mathbf{e}_j$, $i\,\mathbf{e}_i\,\mathbf{e}_j = -\,\varepsilon_{ijk}\,\mathbf{e}_j\,\mathbf{e}_k$, $i\,\mathbf{e}_0\,\mathbf{e}_1\,\mathbf{e}_2\,\mathbf{e}_3 = -\,1$. Furthermore, this complex Clifford algebra is then $\mathbb{C}$-isomorphic to the Clifford algebra of $\mathbb{C}^3$ when given the Euclidian scalar product. Hence, we see that the isomorphism of $SO_0(3, 1)$ with $SO(3; \mathbb{C})$ has a corresponding statement in terms of the relevant Clifford algebras.

All of the foregoing suggests that the most natural representation of the Clifford algebra of Minkowski space for the purposes of electromagnetism is in terms of the complex Clifford algebra $C(\mathbb{C}^3, \delta)$. This, in turn, suggests that one might consider the use of 3-component spin vectors for the field space of wave functions that represent electron/positrons and photons.

For a discussion some of the possible issues associated with generalizing Clifford algebras to projective geometry, see Conradt [**21**, **22**], as well as Hestenes and Ziegler [**23**].

### 4.6 Projective geometry and wave motion

In order to do justice to the topic of this section, we would have to devote at least as much time and detail to developing it, as well as the present topic. Hence, since the present article is growing lengthy as it is, we shall confine ourselves to some general observations about the role of projective geometry in wave motion.

First, we note that the geometric structure of Minkowski space is intrinsically related to the manner by which electromagnetic waves propagate, since the Minkowski quadratic form first appeared in relativity as the symbol of the linear wave equation that one derives from Maxwell's equations and the equation for the light cone is the characteristic equation that is associated with that symbol. Hence, one can justify the assertion that the geometry of spacetime is essentially subordinate to the laws of electromagnetism. In particular, it is the electromagnetic constitutive law of a region of spacetime that dictates its geometry. Hence, the search for the most fundamental laws of spacetime structure should shift from the Riemannian picture of tangent spaces with a Lorentzian structure defined on them to spaces of bivectors or 2-forms with an electromagnetic constitutive law defined on them. This suggests a shift in emphasis from metric geometry to projective geometry.

Since it is widely asserted that it is mechanics, not field theory, that is the most fundamental level of physics, one might also consider what happens to mechanics when one goes from tangent vectors to 2-forms. The answer is simply that rather than considering the motion of points in accordance with dynamical laws that are defined by vector fields – or at least line fields – one must look at the motion of surfaces that are defined by one 2-form and governed by dynamical laws that are defined by another. The surfaces in question are the intersections of the transversal hypersurfaces in spacetime that are defined – perhaps only locally – by one 1-form whose annihilating hyperplanes are tangent to the proper time simultaneity hypersurfaces and another 1-form whose hyperplanes are tangent to the isophase hypersurfaces. The resulting intersection represents points that all have the same value of the wave phase and the same value of proper time; i.e., the *wave fronts* ([10]).

Hence, mechanically the shift from tangent vectors to 2-planes is simply the shift from pointlike mechanics to wave mechanics. Just as physics regards the transition from wave mechanics to point mechanics as a classical limit that is analogous to the transition from wave optics to geometrical optics, one should also regard transition from the electromagnetic constitutive structure on the bundle of 2-forms to the Lorentzian structure on the spacetime tangent bundle as a sort of "classical limit," as well. Whether the methods of geometric asymptotics (see Guillemin and Sternberg [**25**] and Perlick [**26**]) illuminate this transition in the present case is yet to be explored.

Another important topic of wave motion in which projective plays a key role is in the various equivalent formulations of Huygens's principle (see Baker and Copson [**27**] and Arnold [**28**]), one which involves the role of contact transformations in the description of wave motion. Such transformations involve tangent hyperplanes in the same way that point mechanics involves tangent vectors. Hence, they are naturally defined in terms of the projectivized cotangent bundle of spacetime. One cannot help but notice that there are deep geometric – and even topological – subtleties that are associated with the difference between a tangent vector and a 1-form that the traditional index formulation of those objects is largely opaque to.

For a possible application of the methods of projective geometry to wave mechanics and quantum physics, one is encouraged to consider the work of Gschwind [**29-31**].

## 5 Discussion

Perhaps the most compelling direction to pursue at this point was suggested by the discussion of the last section, namely, the representation of wave mechanics itself – both classical and quantum – in terms of geometric structures that pertain to the bundle of 2-forms on the spacetime manifold. The particular case of electromagnetic waves would then be an immediate specialization of the methodology.

A conspicuous limitation of the entire foregoing discussion was the limitation that was imposed by assuming that the electromagnetic constitutive law in question was linear. Although the departure of physical electromagnetic constitutive laws from linearity is most commonly associated with nonlinear optics, in which the optical media can undergo phase transitions in the realm of large electromagnetic field amplitudes, nevertheless, quantum electrodynamics makes such a thorough use of the concept of vacuum polarization that one suspects that nonlinear electromagnetic constitutive laws may be more than empirically applicable to that realm of problems. Indeed, it is reasonable that the "quantum leap" that physics took between the acceptance of Maxwell's linear theory of electromagnetism in vacuo and the eventual development of quantum electrodynamics was largely an attempt to get around the breakdown of Maxwell's linear laws in the realm of large field amplitudes, such as the electric fields of elementary charges – historically, electrons and nuclei – in the realm of small neighborhoods of atomic dimensions. Hence, one should not regard the linearity in Maxwell's theory as suggestive of any first principle of Nature, any more than Hooke's law deserves such a distinction.

[10] For more details on this picture of wave motion see Delphenich [**24**] and the references cited therein.